\documentstyle[12pt]{article}
\input{epsf}
\catcode`\@=11

\@addtoreset{equation}{section}
\def\theequation{\arabic{section}.\arabic{equation}}

\catcode`\@=11

\def\section{\@startsection{section}{1}{\z@}{3.5ex plus 1ex minus
   .2ex}{2.3ex plus .2ex}{\large\bf}}
\def\eqnarray{\let\@currentlabel=\theequation\refstepcounter{equation}
    \global\@eqnswtrue
    \global\@eqcnt\z@\tabskip\@centering\let\\=\@eqncr
    $$\halign to \displaywidth\bgroup\@eqnsel\hskip\@centering
      $\displaystyle\tabskip\z@{##}$&\global\@eqcnt\@ne
       \hfil${{}##{}}$\hfil
      &\global\@eqcnt\tw@ $\displaystyle\tabskip\z@{##}$\hfil
       \tabskip\@centering&\llap{##}\tabskip\z@\cr}
\def\lefteqn#1{\hbox to 4\arraycolsep{$\displaystyle #1$\hss}}
\def\thesection{\arabic{section}.}

\def\appendix{\setcounter{section}{0}
        \def\thesection{Appendix.}
        \def\theequation{\Alph{section}.\arabic{equation}}}

\long\def\@makefntext#1{\parindent 0cm\noindent
\hbox to 1em{\hss$^{\@thefnmark}$}#1}

\def\fnum@figure{\small\figurename~\thefigure}

\def\IR{{\hbox{{\rm I}\kern-.2em\hbox{\rm R}}}}
\def\IH{{\hbox{{\rm I}\kern-.2em\hbox{\rm H}}}}
\def\IC{{\ \hbox{{\rm I}\kern-.6em\hbox{\bf C}}}}
\def\IZ{{\hbox{{\rm Z}\kern-.4em\hbox{\rm Z}}}}
\def\rref#1{(\ref{#1})}
\def\xx{{\vphantom{-1}}}
\def\comp{{\scriptstyle\circ}}
\def\Tr{\hbox{Tr}}
\def\delbar{\,\overline{\mathop{\!\nabla\!}}\,}
\newcommand{\beq}{\begin{equation}}
\newcommand{\eeq}{\end{equation}}
\newcommand{\NPB}[1]{{\sl Nucl.~Phys.}~{\bf B#1}}
\newcommand{\Ann}[1]{{\sl Ann.~Phys.}~{\bf #1}}
\newcommand{\CMP}[1]{{\sl Commun.~Math.~Phys.}~{\bf #1}}
\newcommand{\PLB}[1]{{\sl Phys.~Lett.}~{\bf B#1}}

\newcommand{\PRL}[1]{{\sl Phys.~Rev.~Lett.}~{\bf #1}}
\newcommand{\PTP}[1]{{\sl Prog.~Theor.~Phys.}~{\bf #1}}
\newcommand{\MPLA}[1]{{\sl Mod.~Phys.~Lett.}~{\bf A#1}}
\newcommand{\IJMPA}[1]{{\sl Int.~J.~Mod.~Phys.}~{\bf A#1}}

\newcommand{\IJMPD}[1]{{\sl Int.~J.~Mod.~Phys.}~{\bf D#1}}
\newcommand{\CQG}[1]{{\sl Class.~Quant.~Grav.}~{\bf #1}}
\newcommand{\PRD}[1]{{\sl Phys.~Rev.}~{\bf D#1}}

\newcommand{\JMP}[1]{{\sl J.~Math.~Phys.}~{\bf #1}}
\newcommand{\GRG}[1]{{\sl Gen.~Rel.~Grav.}~{\bf #1}}
\newcommand{\IJTP}[1]{{\sl Int.~J.~Theor.~Phys.}~{\bf #1}}
\begin{document}
%
%
%
%
\def\citen#1{%
\edef\@tempa{\@ignspaftercomma,#1, \@end, }
\edef\@tempa{\expandafter\@ignendcommas\@tempa\@end}%
\if@filesw \immediate \write \@auxout {\string \citation {\@tempa}}\fi
\@tempcntb\m@ne \let\@h@ld\relax \let\@citea\@empty
\@for \@citeb:=\@tempa\do {\@cmpresscites}%
\@h@ld}
%
\def\@ignspaftercomma#1, {\ifx\@end#1\@empty\else
   #1,\expandafter\@ignspaftercomma\fi}
\def\@ignendcommas,#1,\@end{#1}
%
%
\def\@cmpresscites{%
 \expandafter\let \expandafter\@B@citeB \csname b@\@citeb \endcsname
 \ifx\@B@citeB\relax 
    \@h@ld\@citea\@tempcntb\m@ne{\bf ?}%
    \@warning {Citation `\@citeb ' on page \thepage \space undefined}%
 \else
    \@tempcnta\@tempcntb \advance\@tempcnta\@ne
    \setbox\z@\hbox\bgroup 
    \ifnum\z@<0\@B@citeB \relax
       \egroup \@tempcntb\@B@citeB \relax
       \else \egroup \@tempcntb\m@ne \fi
    \ifnum\@tempcnta=\@tempcntb 
       \ifx\@h@ld\relax 
          \edef \@h@ld{\@citea\@B@citeB}%
       \else 
          \edef\@h@ld{\hbox{--}\penalty\@highpenalty \@B@citeB}%
       \fi
    \else   
       \@h@ld \@citea \@B@citeB \let\@h@ld\relax
 \fi\fi%
 \let\@citea\@citepunct
}
%
\def\@citepunct{,\penalty\@highpenalty\hskip.13em plus.1em minus.1em}%
%
%
\def\@citex[#1]#2{\@cite{\citen{#2}}{#1}}%
%
%
\def\@cite#1#2{\leavevmode\unskip
  \ifnum\lastpenalty=\z@ \penalty\@highpenalty \fi 
  \ [{\multiply\@highpenalty 3 #1
      \if@tempswa,\penalty\@highpenalty\ #2\fi 
    }]\spacefactor\@m}
\let\nocitecount\relax  
%
\begin{titlepage}
\vspace{.5in}
\begin{flushright}
UCD-95-6\\
February 1995\\
gr-qc/9503024\\
\end{flushright}
\vspace{.5in}
\begin{center}
{\Large\bf
 Lectures on (2+1)-Dimensional Gravity}\\
\vspace{.4in}
{S.~C{\sc arlip}\footnote{\it email: carlip@dirac.ucdavis.edu}\\
       {\small\it Department of Physics}\\
       {\small\it University of California}\\
       {\small\it Davis, CA 95616}\\{\small\it USA}}
\end{center}

\vspace{.5in}
\begin{center}
{\large\bf Abstract}
\end{center}
\begin{center}
\begin{minipage}{4.75in}
{\small
These lectures briefly review our current understanding of
classical and quantum gravity in three spacetime dimensions,
concentrating on the quantum mechanics of closed universes
and the (2+1)-dimensional black hole.  Three formulations of
the classical theory and three approaches to quantization are
discussed in some detail, and a number of other approaches are
summarized.  An extensive, although by no means complete, list
of references is included.  (Lectures given at the First Seoul
Workshop on Gravity and Cosmology, February 24-25, 1995.)
}
\end{minipage}
\end{center}
\end{titlepage}
\addtocounter{footnote}{-1}

\section{Introduction \label{Intro}}

General relativity is a notoriously difficult theory.  At the
classical level, such fundamental issues as cosmic censorship, the
nature of singularities, and the conditions for formation of
closed timelike curves remain unresolved.  At the quantum level,
the situation is even worse: despite some sixty years of research,
we cannot yet say with any confidence that we understand the
basic conceptual foundations of quantum gravity.

Faced with such difficulties, it is natural to look for simpler
models that share important features with general relativity.  The
choice of model depends on what questions one wishes to ask, but
for many purposes---especially in the realm of quantum gravity---a
particularly useful model is general relativity in three spacetime
dimensions.  Work on (2+1)-dimensional gravity dates back at least
to 1963 \cite{Staru}, and occasional articles appeared over the next
twenty years\cite{Leut,Collas,Clem1}.  But credit for the recent growth of
interest should probably go to two groups: Deser, Jackiw, and 't~Hooft
\cite{DJtH,DJ,tH}, who examined the classical and quantum dynamics
of point sources, and Witten \cite{Wita,Witb,Witc}, who rediscovered
and explored the representation of (2+1)-dimensional gravity as a
Chern-Simons theory.\footnote{The Chern-Simons representation was
first pointed out, I believe, by Ach\'ucarro and Townsend \cite{Achu}.}

Hundreds of papers about (2+1)-dimensional gravity have appeared since
the pioneering work of the mid-1980's, and the subject has become far
too extensive for me to summarize fairly in these lectures.  I have
therefore chosen to focus on a few key aspects: classical gravity for
empty, spatially compact universes, a few approaches to the quantization
of this classical physics (``(2+1)-dimensional quantum cosmology''),
and the classical and quantum mechanics of the (2+1)-dimensional black
hole.  I am necessarily leaving out many interesting topics: classical
point sources and closed timelike curves (see, for instance, \cite{Gott,
Gott2,tH2,Mena}); the classical and quantum behavior of point particles
(for a partial sample of extensive work, see \cite{DJtH,DJ,tH,Jackiw,
Rocek,Clem,Gerbert,SC1,Koehler,SCm,Vaz,Mansouri,Ortiz}); matter couplings
(see, for instance, \cite{Geg,CarGeg,Mizo,AshVar,Pel,Bona}); supergravity
(see, for instance, \cite{Achu,Dayi,Ko2,deWit,Nicolai}; asymptotic behavior
and global ``charges'' \cite{Henn,Deser,Brown}; the radial gauge
\cite{Menotti,Menotti2}; topologically massive gravity \cite{DJTemp,
DesY,Kleppe}; Poincar\'e gauge theory \cite{Grig1,Grig2,Kawai}; the
construction of observables from topological field theories \cite{Brooks,
Brooks2}; and an assortment of other subjects.

The structure of these lectures is as follows.  In the next section,
I discuss the field equations of classical general relativity in
2+1 dimensions, and obtain the set of solutions for spacetimes with
the topology $\IR\!\times\!\Sigma$, where $\Sigma$ is a closed surface
of genus $g\!>\!0$.  I summarize three classically equivalent descriptions
of these solutions, in terms of gluing patterns of geometric structures,
gauge field holonomies arising from a Chern-Simons formalism, and
conventional metric (ADM) variables.  I next provide a somewhat more
detailed analysis of the torus universe (subsection \ref{Tor}) and
the (2+1)-dimensional black hole (subsection \ref{BH}).  In section 3,
I use these classical results to formulate and compare three inequivalent
approaches to quantization, again giving details for the $\IR\!\times\!T^2$
topology, and I briefly summarize several other approaches to quantization
(subsection \ref{Other}).  In section 4, I address a few remaining
topics, including black hole thermodynamics and topology-changing
amplitudes.

\section{Classical Gravity in 2+1 Dimensions \label{Clas}}
\setcounter{footnote}{0}

The goal of this section is to give a description of the space of classical
solutions of general relativity in 2+1 dimensions.  In fact, I shall
derive three different---although classically equivalent---descriptions,
coming from a direct analysis of the geometry (subsection \ref{GS}), a
first-order ``gauge'' formalism (subsection \ref{CS}), and a more
traditional metric formalism (subsection \ref{ADM}).

Let us begin by examining the reasons for the simplicity of general
relativity in 2+1 dimensions.  In any spacetime, the curvature tensor
may be decomposed into a curvature scalar $R$, a Ricci tensor
$R_{\mu\nu}$, and a remaining trace-free, conformally invariant
piece, the Weyl tensor $C_{\mu\nu\rho}{}^{\sigma}$.  In 2+1 dimensions,
however, the Weyl tensor vanishes identically, and the full curvature
tensor is determined algebraically by the curvature scalar and the
Ricci tensor:
\beq
R_{\mu\nu\rho\sigma} = g_{\mu\rho}R_{\nu\sigma}
+ g_{\nu\sigma}R_{\mu\rho} - g_{\nu\rho}R_{\mu\sigma}
- g_{\mu\sigma}R_{\nu\rho} - {1\over2}
(g_{\mu\rho}g_{\nu\sigma} - g_{\mu\sigma}g_{\nu\rho})R .
\label{a1}
\eeq
In particular, this implies that any solution of the vacuum Einstein
field equations is {\em flat}, and that any solution of the field
equations with a cosmological constant,
\beq
R_{\mu\nu} = 2\Lambda g_{\mu\nu} ,
\label{a2}
\eeq
has constant curvature.  Physically, a (2+1)-dimensional spacetime has
no local degrees of freedom: there are no gravitational waves in the
classical theory, and no gravitons in the quantum theory.

Our aim was to find a simple model with which to explore conceptual
issues of general relativity.  At first sight, though, this model seems
{\em too\/} simple.  Indeed, most of us were taught in our first course
in general relativity that the vanishing of $R_{\mu\nu\rho\sigma}$ implies
that the metric is just the ordinary Minkowski metric $\eta_{\mu\nu}$.
Certainly, quantum gravity will be simple if there is no gravity, but
it won't teach us much.

Fortunately, life is a bit more complicated.  The vanishing of the
curvature tensor means that any point in a spacetime $M$ has a
neighborhood that is isometric to Minkowski space $(V^{2,1},\eta)$.  If
$M$ has a trivial topology, a single neighborhood can be extended globally,
and the geometry is indeed trivial; but if $M$ contains noncontractible
curves, such an extension may not be possible.  The existence of
``global geometry'' for spacetimes with nontrivial topologies leads
us to our first description of the solutions of the field equations
\rref{a2}: the description in terms of geometric structures.

\subsection{Geometric Structures \label{GS}}

To understand the distinction between local and global geometry,
let us start with the simpler case of a flat two-dimensional torus
$T^2$.  (This example will be important later.)  Any such torus can be
described as a parallelogram in the complex plane with opposite sides
identified, and up to an overall rescaling, the vertices of such a
parallelogram may be placed at the points $0$, $1$, $\tau$, and $\tau+1$,
where the modulus $\tau = \tau_1+i\tau_2$ is an arbitrary complex number
with positive imaginary part (see figure 1).  The identification of the
sides is an isometry of the flat metric on the plane, so $T^2$ inherits
a flat metric.  But it is easy to see that tori with different values of
$\tau$ are not, in general, isometric.\footnote{Certain valus of $\tau$
{\em are\/} actually connected by ``large'' diffeomorphisms; we shall
return to this issue in subsection \ref{MCG}.}  In other words, the
requirement that $R_{\mu\nu\rho\sigma}$ vanish determines the local
geometry, but nevertheless leaves us with a two-parameter family of
globally inequivalent geometries.

\begin{figure}
\begin{center}
\vspace{-2ex}
\leavevmode
\epsfysize=1.75in \epsfbox{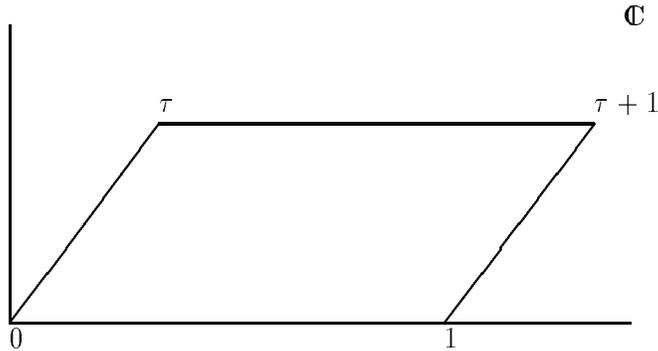}
\end{center}
\vspace{-3ex}
\caption{\small A  flat torus of modulus $\tau$ is represented
as a parallelogram with opposite sides identified.}
\end{figure}

This ``gluing'' process can alternatively be described as a quotient
space construction.  Consider the group $G_\tau$ of isometries of the
complex plane generated by the translations
\begin{eqnarray}
z&\mapsto& z+1 \nonumber\\
z&\mapsto& z+\tau .
\label{a2a}
\end{eqnarray}
This group acts properly discontinuously on the plane, and the quotient
space $\IC/G_\tau$ is precisely the flat torus with modulus $\tau$.  A
similar quotient construction exists for closed surfaces of genus $g\!>\!1$
\cite{Abikoff,Fish,Keen}.  By the uniformization theorem, any such surface
$\Sigma$ admits a metric of constant curvature $-1$, which can be lifted
to the universal covering space of $\Sigma$, the hyperbolic plane $\IH^2$.
Conversely, $\Sigma$ can be recovered as a quotient $\IH^2/G$, where $G$
is a so-called Fuchsian group, a discrete subgroup of $\hbox{PSL}(2,\IR)$.
($\hbox{PSL}(2,\IR)$ is the group of isometries of the constant negative
curvature metric on $\IH^2$.)

Let us now apply a similar analysis to a flat (2+1)-dimensional
spacetime.  The vanishing of the curvature implies that $M$ can be
covered by a set of contractible coordinate patches $U_i$, each isometric
to Minkowski space $V^{2,1}$ with the standard Minkowski metric
$\eta_{\mu\nu}$.  In general, though, these patches must be ``glued
together'' by transition functions $\gamma_{ij}$ on the intersections
$U_i\cap U_j$, which determine how points are identified.  Since the
metrics on $U_i$ and $U_j$ are identical, these transition functions
must be isometries of $\eta_{\mu\nu}$, that is, elements of the Poincar\'e
group ISO(2,1).  As in the case of the flat torus, the global geometry is
hidden in these identifications.

This construction is an example of what Thurston calls a geometric
structure \cite{Thur,Canary,Gold3,SullThur}, in this case a Lorentzian
or (ISO(2,1),$V^{2,1}$) structure.  In general, a $(G,X)$ manifold is
one that is locally modeled on $X$, just as an ordinary $n$-dimensional
manifold is modeled on $\IR^n$.  More precisely, let $G$ be a Lie group
that acts analytically on some $n$-manifold $X$, the model space, and
let $M$ be another $n$-manifold.  A $(G,X)$ structure on $M$ is a set
of coordinate patches $U_i$ covering $M$ with ``coordinates'' $\phi_i\!:
U_i\!\rightarrow\!X$ taking their values in the model space and with
transition functions $\gamma_{ij}^\xx = \phi_i^\xx\comp\phi_j^{\ -1}|
U_i\cap U_j$ in $G$.  While this general formulation is not very widely
known among physicists, specific examples are familiar.  The flat torus
is one; a general Riemann surface is another, since the uniformization
theorem guarantees that any surface of genus $g\!>\!1$ admits a hyperbolic
(that is, $(\IH^2,\hbox{PSL(2,$\IR$)})$) structure.

A fundamental invariant of a $(G,X)$ structure is its holonomy group,
which can be thought of as a measure of the failure of a single coordinate
patch to extend around a closed curve.  Let $M$ be a $(G,X)$ manifold
containing a closed path $\gamma$.  We can cover $\gamma$ with coordinate
charts
\beq
\phi_i: U_i\rightarrow X,\qquad i=1,\dots,n
\label{a3}
\eeq
with constant transition functions $g_i\in G$ between $U_i$ and $U_{i+1}$,
i.e.,
\begin{eqnarray}
\phi_i|U_i\cap U_{i+1} &=& g_i\comp \phi_{i+1}|U_i\cap U_{i+1}\nonumber\\
\phi_n|U_n\cap U_{1} &=& g_n\comp \phi_{1}|U_n\cap U_{1}
\label{a4}
\end{eqnarray}
(see figure 2).
\begin{figure}
\begin{center}
\leavevmode
\epsfysize=2.9in \epsfbox{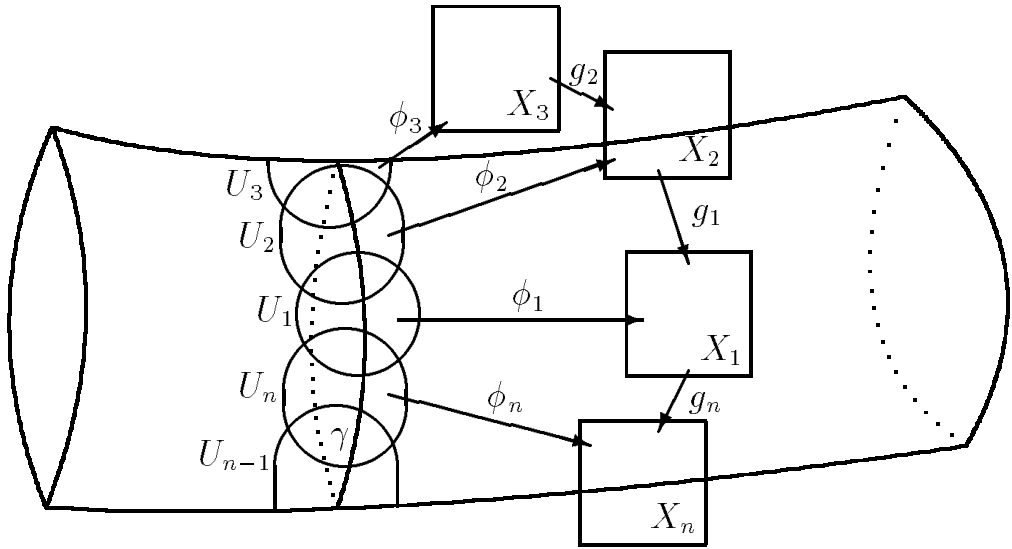}
\end{center}
\vspace{-2ex}
\caption{\small The curve $\gamma$ is covered by coordinate patches $U_i$,
with transition functions $g_i\!\in\!G$.  The composition $g_1\comp\dots
\comp g_n$ is the holonomy of the curve.}
\end{figure}
Let us now try to analytically continue the coordinate $\phi_1$ from the
patch $U_1$ to all of $\gamma$.  We begin with a coordinate transformation
in $U_2$ that replaces $\phi_2$ by ${\phi_2}'=g_1\comp\phi_2$, thus extending
$\phi_1$ to $U_1\cup U_2$.  Continuing this process along $\gamma$, we will
eventually reach the final patch $U_n$, which again overlaps $U_1$.  If the
new coordinate function ${\phi_n}'=g_1\comp\dots\comp g_{n-1}\comp\phi_n$
happens to agree with $\phi_1$ on $U_n\cap U_1$, we will have succeeded in
covering $\gamma$ with a single patch.  Otherwise, the holonomy $\rho$,
defined as
\beq
\rho(\gamma) = g_1\comp\dots\comp g_n \in G,
\label{a5}
\eeq
measures the obstruction to such a covering.

It may be shown that the holonomy of a curve $\gamma$ depends only on
its homotopy class \cite{Thur}.  In fact, the holonomy defines a group
homomorphism
\beq
\rho: \pi_1(M)\rightarrow G .
\label{a6}
\eeq
The homomorphism $\rho$ is not quite uniquely determined by the geometric
structure, but it is unique up to conjugation by a constant element
$h\!\in\!G$, i.e., $\rho\!\mapsto\!h\cdot\rho\cdot h^{-1}$ \cite{Thur}.
For the case of (2+1)-dimensional gravity, where $G$ is the Poincar\'e
group, we thus obtain a space of holonomies of the form
\begin{eqnarray}
{\cal M} &=& \hbox{Hom}(\pi_1(M),\hbox{ISO(2,1)})/\sim ,\nonumber\\
\rho_1&\sim&\rho_2 \ \ \hbox{if}\ \ \rho_2 = h\cdot\rho_1\cdot h^{-1},
\quad h\in\hbox{ISO(2,1)} .
\label{a7}
\end{eqnarray}

We shall later make use of the fact (see, for instance, \cite{SCm} or
\cite{SC2}) that $\cal M$ has the structure of a cotangent bundle,
${\cal M}\!\approx\!T^*{\cal N}$, where $\cal N$ is the space of
SO(2,1) projections of the ISO(2,1) holonomies,
\begin{eqnarray}
{\cal N} &=& \hbox{Hom}(\pi_1(M),\hbox{SO(2,1)})/\sim ,\nonumber\\
\hat\rho_1&\sim&\hat\rho_2 \ \ \hbox{if}\ \ \hat\rho_2 =
\hat h\cdot\hat\rho_1\cdot \hat h^{-1},
\quad \hat h\in\hbox{SO(2,1)} .
\label{a7a}
\end{eqnarray}
(I will use the symbol $\hat\rho$ to denote holonomies in $\cal N$,
that is, SO(2,1)-valued holonomies, reserving $\rho$ to denote
ISO(2,1)-valued holonomies in $\cal M$.)

A flat spacetime geometry thus determines a holonomy $\rho\!\in\!{\cal M}$.
We can now ask whether, conversely, such a holonomy uniquely determines
a geometry.  In other words, have we succeeded in completely characterizing
the solutions of the vacuum Einstein field equations in 2+1 dimensions?

For a general spacetime $M$, the answer to this question is not known.
However, Mess has studied this question for the case of spacetimes with
topologies of the form $\IR\!\times\!\Sigma$, where $\Sigma$ is a closed
surface \cite{Mess}.  He shows that the holonomy group determines a
unique ``maximal'' spacetime $M$---to be precise, a  spacetime constructed
as a domain of dependence of a spacelike surface $\Sigma$.  Mess also
demonstrates that the holonomy group acts properly discontinuously on a
region $W\!\subset\!V^{2,1}$ of Minkowski space, and that $M$ can be
obtained as the quotient space $W/\langle\rho[\gamma]\rangle$, thus
generalizing the quotient construction for the flat torus we considered
above.  As we shall see later, this construction can be a powerful tool
for obtaining a description of $M$ in reasonably standard coordinates,
for instance in a time slicing by surfaces of constant mean curvature.

Note that if $M\!\approx\!\IR\!\times\!\Sigma$, the fundamental group
$\pi_1(M)$ is isomorphic to $\pi_1(\Sigma)$.  The holonomies
$\rho\!\in\!{\cal M}$ and $\hat\rho\!\in\!{\cal N}$ may thus be
determined from data on a single spatial slice $\Sigma$.  This is a
first indication of the ``frozen time'' problem that will be a central
issue in subsection \ref{CCQ}.

If the cosmological constant $\Lambda$ is nonzero, a similar
construction is possible.  Our (2+1)-dimensional spacetime now has
constant curvature, and the coordinate patches $U_i$ will be
isometric to de Sitter or anti-de Sitter space.  The gluing isometries
correspondingly become elements of $\hbox{SO}(2,2)$ (for $\Lambda\!<\!0$)
or $\hbox{SO}(3,1)$ (for $\Lambda\!>\!0$), and the holonomies are now
elements of one of these groups.  A simple example of such a construction
is given by Fujiwara \cite{Fujiwara}.

For topologies $\IR\!\times\!\Sigma$ with $\Lambda<0$, Mess has shown
that the holonomy group again determines a unique maximal spacetime.  If
$\Lambda\!>\!0$, this is no longer true: a given holonomy group determines
an infinite family of nonisometric spacetimes.  For the simplest nontrivial
topology, $M\!\approx\!\IR\!\times\!T^2$, Ezawa has explicitly obtained
the set of geometries that arise from a given holonomy group \cite{Ezawa};
see \cite{Witc} for some speculation about the physical significance
of this redundancy.

\subsection{The Chern-Simons Formulation \label{CS}}

The method of geometric structures gives us an explicit solution of the
Einstein field equations in 2+1 dimensions, but it is a rather unusual
one.  In particular, we have not had to solve a single differential
equation.  To make contact with more conventional results, let us now
consider an alternative approach, starting from the first-order form
of the Einstein action.  (For a review of the first- and second-order
formalism, see \cite{Rom}.)

The fundamental variables are now a triad $e_\mu{}^a$---technically,
a section of the bundle of orthonormal frames---and a spin connection
$\omega_\mu{}^a{}_b$.  The Einstein-Hilbert action can be written as
\beq
I_{\hbox{\scriptsize grav}} = 2\int_M\,e^a\wedge
  \left(d\omega_a + {1\over2}\epsilon_{abc}\omega^b\wedge\omega^c\right) ,
\label{a8}
\eeq
where $e^a = e_\mu{}^a dx^\mu$ and $\omega^a = {1\over2}\epsilon^{abc}
\omega_{\mu bc}dx^\mu$.  (My units are such that $16\pi G = 1$.)  The
action is invariant under local SO(2,1) transformations,
\begin{eqnarray}
\delta e^a &=& \epsilon^{abc}e_b\tau_c \nonumber\\
\delta \omega^a &=& d\tau^a + \epsilon^{abc}\omega_b\tau_c ,
\label{a9}
\end{eqnarray}
as well as ``local translations,''
\begin{eqnarray}
\delta e^a &=& d\sigma^a + \epsilon^{abc}\omega_b\sigma_c \nonumber\\
\delta \omega^a &=& 0 .
\label{a10}
\end{eqnarray}
$I_{\hbox{\scriptsize grav}}$ is also invariant under diffeomorphisms of
$M$, of course, but this is not an independent symmetry: Witten has shown
that when the triad $e_\mu{}^a$ is invertible, diffeomorphisms in the
connected component of the identity are equivalent to transformations
of the form \rref{a9}--\rref{a10} \cite{Wita}.  An explicit construction
of the generators of diffeomorphisms in terms of generators of gauge
transformations has been carried out by Ba\~nados \cite{Ban}; see also
\cite{Rom}.

The equations of motion coming from the action \rref{a8} are easily
derived:
\beq
T^a[e,\omega] = de^a + \epsilon^{abc}\omega_b\wedge e_c = 0
\label{a11}
\eeq
and
\beq
R^a[\omega]
  = d\omega^a + {1\over2}\epsilon^{abc}\omega_b\wedge\omega_c = 0 .
\label{a12}
\eeq
The first of these is the standard torsion-free condition that determines
$\omega$ in terms of $e$.  The second then implies that the connection
$\omega$ is flat, or equivalently that the curvature of the metric
$g_{\mu\nu} = e_\mu{}^a e_\nu{}^b \eta_{ab}$ vanishes, thus reproducing
the field equations of the last subsection.  In this formulation, the
significance of the global geometry is clear: if $M$ is topologically
nontrivial, a flat connection can still give rise to nonvanishing
Aharonov-Bohm phases around noncontractible curves.

There are several ways to understand the solutions of equations
\rref{a11}--\rref{a12}.  The easiest is to note that the flat connection
$\omega$ is determined by its holonomies,\footnote{Note that the meaning
of the term ``holonomy'' has changed here---the holonomy of a flat
connection is determined by parallel transport in a fiber bundle, not
by the gluing of coordinate patches.} that is, by a homomorphism
$\hat\rho\!\in\!{\cal N}$, where $\cal N$ is the same space of
homomorphisms that appeared in equation \rref{a7a} of the last subsection.
These holonomies are just the Wilson loops of the connection,
\beq
\hat\rho[\gamma] = P\exp\left\{\int_\gamma\omega^a{\cal J}_a\right\} ,
\label{a12a}
\eeq
where $P$ denotes path ordering and the ${\cal J}_a$ are the generators
of the gauge group SO(2,1).  Note that the flatness of $\omega$ implies
that $\hat\rho[\gamma]$ depends only on the homotopy class of $\gamma$.
Moreover, equation \rref{a11} implies that $e$ is a cotangent vector to
the space of flat connections.  Indeed, if $\omega(s)$ is a curve in the
space of flat connections, the derivative of \rref{a12} with respect to
$s$ gives
\beq
d\left({d\omega^a\over ds}\right) +
 \epsilon^{abc}\omega_b\wedge\left({d\omega_c\over ds}\right) = 0 ,
\label{a13}
\eeq
which can be identified with \rref{a11} with
\beq
e^a = {d\omega^a\over ds} .
\label{a14}
\eeq
A solution $(\omega,e)$ of the first-order field equations is thus
labeled by a point in $T^*{\cal N}$, just as in the geometric structure
approach.  This is not a coincidence:  given a $(G,X)$ structure on a
manifold $M$, there is a standard direct construction of a corresponding
flat connection, as discussed in references \cite{Gold3} and \cite{SC2}.

We can learn more about the field equations \rref{a11}--\rref{a12} by
observing that the one-forms $e^a$ and $\omega^a$ can be combined to
form a single ISO(2,1) connection \cite{Wita,Achu}.  The Lie algebra
of ISO(2,1) has generators ${\cal J}^a$ and ${\cal P}^b$, with commutation
relations
\beq
\left[{\cal J}^a, {\cal J}^b\right] = \epsilon^{abc}{\cal J}_c ,\qquad
\left[{\cal J}^a, {\cal P}^b\right] = \epsilon^{abc}{\cal P}_c ,\qquad
\left[{\cal P}^a, {\cal P}^b\right] = 0 .
\label{a15}
\eeq
If we write a single connection one-form
\beq
A = e^a{\cal P}_a + \omega^a{\cal J}_a
\label{a16}
\eeq
and define a ``trace,'' an invariant inner product on the Lie algebra, by
\beq
\Tr\left({\cal J}^a{\cal P}^b\right) = \eta^{ab} , \qquad
\Tr\left({\cal J}^a{\cal J}^b\right) = \Tr\left({\cal P}^a{\cal P}^b\right)
= 0 ,
\label{a17}
\eeq
it is easy to check that the first-order action \rref{a1} is simply the
Chern-Simons action \cite{WitJones} for $A$,
\beq
I_{\hbox{\scriptsize CS}} = {k\over4\pi}\int_M\,\Tr
  \left\{ A\wedge dA + {2\over3}A\wedge A\wedge A\right\} ,
\label{a18}
\eeq
where $k=4\pi$ with my choice of units.  Furthermore, the gauge
transformations \rref{a9}--\rref{a10} may now be reinterpreted as
standard ISO(2,1) gauge transformations of $A$.

Now, the field equations of a Chern-Simons theory simply require
that $A$ be flat \cite{WitJones}.  We thus expect solutions to
the field equations to be labeled by ISO(2,1) holonomies, that is,
homomorphisms $\rho\!\in\!{\cal M}$, where $\cal M$ is  precisely
the space \rref{a7} that appeared in our earlier analysis of
geometric structures.  The equivalence relation in \rref{a7} is now
easy to understand: a gauge transformation $g\!: M\!\rightarrow\!G$
acts on Wilson loops based at $x_0$ by conjugation by $g(x_0)$, so
the quotient in \rref{a7} is simply an expression of gauge invariance.

Note that the traces $\Tr\rho[\gamma]$ are automatically invariant under
conjugation, and thus provide a set of gauge-invariant observables.  In
general, these observables form an overcomplete set;  Nelson and Regge
\cite{NR1,NR3,NR2} and Martin \cite{Martin} have investigated the
identities among them, and Loll has recently proposed a complete subset
of traces \cite{Loll}.

A similar construction is possible when $\Lambda\!\ne\!0$
\cite{Wita,Achu,NRZ}.  For $\Lambda\!=\!-1/\ell^2\!<\!0$, the two SO(2,1)
connections
\beq
A^{(\pm)a} = \omega^a \pm {1\over\ell} e^a
\label{a19}
\eeq
can be treated as independent variables, and the Einstein-Hilbert action
becomes
\beq
I_{\hbox{\scriptsize grav}} =
  I_{\hbox{\scriptsize CS}}[A^{(+)}] -I_{\hbox{\scriptsize CS}}[A^{(-)}] ,
\label{a20}
\eeq
where now $k = \ell\sqrt{2}/8G$ in the conventions of \cite{SC3}. (The
numerical value of $k$ depends on the choice of representation and the
definition of the trace in \rref{a18}.)  For $\Lambda\!>\!0$, the
Einstein-Hilbert action is equivalent to the Chern-Simons action for
the $\hbox{SL}(2,\IC)$ connection
\beq
\tilde A^a = \omega^a + {i\sqrt{\Lambda}} e^a .
\label{a21}
\eeq
For either sign of $\Lambda$, the holonomies of the gauge field reproduce
the holonomies of the corresponding geometric structure discussed in the
last subsection.

As a prerequisite for quantizing these models, we shall need the classical
Poisson brackets among the physical variables.  These are not at all
obvious in the geometric structure approach, but their derivation is
reasonably straightforward in the Chern-Simons picture.  Note first that
the brackets of $e_i{}^a$ and $\omega_j{}^b$ on a slice of constant time
can be read off directly from the action \rref{a8}:
\beq
\{ e_i{}^a(x), \omega_j{}^b(x') \}
  = {1\over2} \eta^{ab}\epsilon_{ij}\delta^2(x-x') .
\label{a22}
\eeq
These brackets, in turn, induce Poisson brackets among the traces of
holonomies that parametrize the space of solutions.  The resulting
Poisson algebra is easiest to work out when $\Lambda\ne0$.  The resulting
brackets were calculated by hand for the genus 1 and genus 2 cases and
then generalized and quantized by Nelson and Regge in reference \cite{NR2}.
Their general form is too complicated to write down here, but the special
case of the torus will be described below.  It is interesting to note that
these brackets are closely related to the symplectic structure on the
abstract space of loops on $\Sigma$ first discovered by Goldman
\cite{Goldman4,GoldHam}.

\subsection{The ADM Formalism \label{ADM}}

While the Chern-Simons formalism described above is fairly close to
the standard first-order description of general relativity in 3+1
dimensions, it is still rather far from the usual metric description.
As Moncrief \cite{Mon} and Hosoya and Nakao \cite{HosNak} have shown,
the metric formalism can also be used to give a full description of
the solutions of the vacuum field equations, at least for spacetimes
with the topology $\IR\!\times\!\Sigma$.

To obtain this description, let us consider an Arnowitt-Deser-Misner
decomposition of the metric $g_{\mu\nu}$,
\beq
ds^2 = N^2dt^2 - g_{ij}(dx^i + N^i dt)(dx^j + N^j dt) ,
\label{a23}
\eeq
for which the action takes the usual form\footnote{In this subsection I
use standard ADM notation: $g_{ij}$ and $R$ refer to the induced metric
and scalar curvature of a time slice, while the spacetime metric and
curvature are denoted ${}^{\scriptscriptstyle(3)}\!g_{\mu\nu}$ and
${}^{\scriptscriptstyle(3)}\!R$.}

\beq
I_{\hbox{\scriptsize grav}}
  = \int\!d^3x \sqrt{-{}^{\scriptscriptstyle(3)}\!g}\>
  ({}^{\scriptscriptstyle(3)}\!R - 2\Lambda)
  = \int dt\int\nolimits_\Sigma d^2x \bigl(\pi^{ij}{\dot g}_{ij}
               - N^i{\cal H}_i -N{\cal H}\bigr) .
\label{a24}
\eeq
Here the canonical momentum $\pi^{ij}$ is given by
\beq
\pi^{ij} = \sqrt{g}\,(K^{ij}- g^{ij}K),
\label{a24a}
\eeq
where $K^{ij}$ is the extrinsic curvature of the surface $t={\rm const.}$,
and the momentum and Hamiltonian constraints in 2+1 dimensions are
\beq
{\cal H}_i = -2\nabla_j\pi^j_{\ i}\,, \qquad
{\cal H} = {1\over\sqrt{g}}\,g_{ij}g_{kl}(\pi^{ik}\pi^{jl}-\pi^{ij}\pi^{kl})
                 -\sqrt{g}(R - 2\Lambda) .
\label{a25}
\eeq

To solve the constraints, we choose the York time slicing \cite{York},
in which the mean (extrinsic) curvature is used as a time coordinate,
$-K=\pi/\sqrt{g}=T$.  In reference \cite{Mon}, Moncrief shows that this
is a good global coordinate choice for classical solutions of the field
equations.  We next select a useful parametrization of the spatial metric
and momentum.  Up to a diffeomorphism, any two-metric on $\Sigma$ can be
written in the form \cite{Abikoff,Fish}
\beq
g_{ij} = e^{2\lambda}{\bar g}_{ij}(m_\alpha) ,
\label{a26}
\eeq
where the $\bar g_{ij}(m_\alpha)$ are a finite-dimensional family of
metrics of constant curvature, labeled by a set of moduli $m_\alpha$.
For the torus, for instance, we can write $\bar g_{ij}=\bar g_{ij}(\tau)$,
where $\tau$ is the modulus introduced at the beginning of subsection
\ref{GS}, corresponding to a spatial metric
\beq
d\sigma^2 = \left| dx + \tau dy\right|^2
\label{a26a}
\eeq
(where $x$ and $y$ have period $1$).  Similarly, any closed surface
of genus $g\!>\!1$ admits a hyperbolic structure (that is, an
$(\IH^2,\hbox{PSL(2,$\IR$)})$ geometric structure), and the corresponding
identifications on $\IH^2$ are labeled by $6g-6$ parameters $m_\alpha$.

The corresponding decomposition of the $\pi^{ij}$ takes the form
\beq
\pi^{ij} = e^{-2\lambda}\sqrt{\bar g} \left(
  p^{ij}+ {1\over2}\bar g^{ij}\pi/\sqrt{\bar g}
  + \delbar^iY^j + \delbar^jY^i - \bar g^{ij}\delbar_kY^k \right),
\label{a27}
\eeq
where $\delbar_i$ is the covariant derivative for the connection
compatible with $\bar g_{ij}$, indices are now raised and lowered with
$\bar g_{ij}$, and $p^{ij}$ is a transverse traceless tensor with respect
to $\delbar_i$, i.e., $\delbar_i\, p^{ij} = 0$.  In the language of
Riemann surfaces, $p^{ij}$ is a holomorphic quadratic differential; the
space of such differentials parametrizes the cotangent space of the moduli
space \cite{Abikoff}.  Roughly speaking, the $p^{ij}$ are conjugate to
the moduli, $\pi$ is conjugate to the scale factor $\lambda$, and the
$Y^i$ are conjugate to spatial diffeomorphisms.

The momentum constraints ${\cal H}_i=0$ now imply that $Y^i=0$, while
the Hamiltonian constraint,
\beq
{\cal H} = -{1\over2}\sqrt{\bar g}e^{2\lambda}(T^2 - 4\Lambda)
  + \sqrt{\bar g}e^{-2\lambda} p^{ij}p_{ij} + 2\sqrt{\bar g}\left[
  \bar\Delta\lambda - {1\over2}\bar R \right] = 0 ,
\label{a28}
\eeq
uniquely determines $\lambda$ as a function of $\bar g_{ij}$ and
$p^{ij}$ \cite{Mon}.  The action \rref{a24} reduces to
\beq
I_{\hbox{\scriptsize grav}}
  = \int dT \left( p^\alpha {dm_\alpha\over dT} - H(m,p,T) \right) ,
\label{a29}
\eeq
where the Hamiltonian is
\beq
H = \int_\Sigma \sqrt g\,d^2x
  = \int_\Sigma e^{2\lambda(m,p,T)}\sqrt{\bar g}\,d^2x
\label{a29a}
\eeq
with $\lambda(m,p,T)$ determined by \rref{a28}, and the $p^\alpha$ are
momenta conjugate to the moduli, given by
\beq
p^\alpha = \int_\Sigma d^2x\,
  p^{ij}{\partial\ \ \over\partial m_\alpha}{\bar g}_{ij} .
\label{a30}
\eeq
The classical Poisson brackets can be read off directly from \rref{a29}:
\beq
\{m_\alpha, p^\beta\} = \delta_\alpha^\beta , \quad
\{m_\alpha, m_\beta\} = \{p^\alpha, p^\beta\} = 0 .
\label{a30a}
\eeq

Three-dimensional gravity is thus reduced once again to a
finite-dimensional system, albeit one with a complicated, time-dependent
Hamiltonian.  It is evident from \rref{a29} that the physical phase
space---and hence the space of solutions---is parametrized by
$(m_\alpha,p^\beta)$, which may be viewed as coordinates for the
cotangent bundle of the moduli space of $\Sigma$.  But mathematicians
have known for some time that the moduli space of a Riemann surface
is homeomorphic\footnote{Strictly speaking, $\cal N$ has a number of
connected components, one of which is homeomorphic to moduli space.}
to the space of holonomies $\cal N$ defined by equation \rref{a7a}
\cite{Macbeath,Goldman2,Goldman3}.  We thus recover the description of the
space of solutions as the cotangent bundle $T^*{\cal N}$.  For the
simple case of a torus universe, $M\approx\IR\!\times\!T^2$, the exact
relationship between the coordinates $(m_\alpha,p^\beta)$ and the
holonomies $\rho[\gamma]$ will be described below.

\subsection{The Mapping Class Group \label{MCG}}

In the presentation so far, I have avoided discussing an important
symmetry of general relativity on topologically nontrivial spacetimes.
The description of a solution of the Einstein field equations in terms
of holonomies (subsections \ref{GS} and \ref{CS}) or moduli and their
conjugate momenta (subsection \ref{ADM}) is invariant under infinitesimal
diffeomorphisms, and therefore under ``small'' diffeomorphisms, those
which can be smoothly deformed to the identity.  But if $M$ is
topologically nontrivial, its group of diffeomorphisms may not be
connected: $M$ may admit ``large'' diffeomorphisms, which cannot be built
up smoothly from infinitesimal deformations.   The group of such large
diffeomorphisms of $M$ (modulo small diffeomorphisms), ${\cal D}(M)$,
is called the mapping class group of $M$; for the torus $T^2$, it is
also known as the modular group.

The archetype of a large diffeomorphism is a Dehn twist of a torus,
which may be described as the operation of cutting $T^2$ along a
circumference to obtain a cylinder, twisting one end of the cylinder
by $2\pi$, and regluing (see figure 3).  Similar transformations
exist for an arbitrary closed surface $\Sigma$, and in fact the Dehn
twists around generators of $\pi_1(\Sigma)$ generate ${\cal D}(\Sigma)$
\cite{Birman,Birman2}.
\begin{figure}
\begin{center}
\leavevmode
\epsfysize=2in \epsfbox{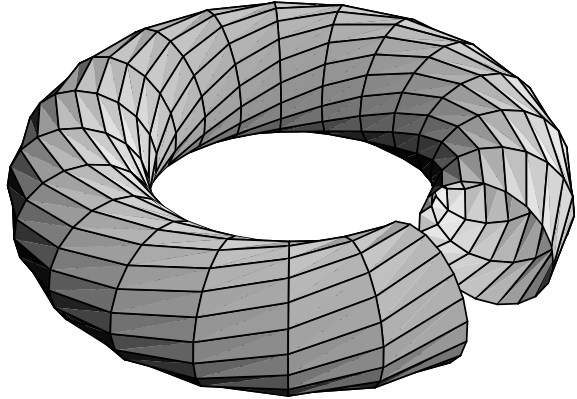}
\end{center}
\vspace{-2ex}
\caption{\small A Dehn twist of a torus is obtained by cutting along one
of the circumferences, rotating one end by $2\pi$, and regluing.}
\end{figure}

It is easy to see that the mapping class group of a spacetime $M$
acts on $\pi_1(M)$, and therefore on the holonomies of subsection \ref{GS}.
As a group of diffeomorphisms, the group also acts on the constant
curvature metrics $\bar g_{ij}$, and hence on the moduli $m_\alpha$ of
subsection \ref{ADM}.  Classically, geometries that differ by actions of
${\cal D}(M)$ are completely equivalent, so the ``true'' space of
solutions for a spacetime $\IR\!\times\!\Sigma$ is really
\beq
T^*{\cal N}(\Sigma)/{\cal D}(\Sigma) .
\label{a31}
\eeq
Quantum mechanically, this equivalence may be relaxed, but wave
functions should at least transform under some unitary representation
of the mapping class group.

\subsection{The Torus Universe \label{Tor}}

The discussion so far has been rather abstract.  For a concrete example,
let us consider the torus universe, $M\!\approx\!\IR\!\times\!T^2$, with a
negative cosmological constant $\Lambda\!=\!-1/\ell^2$ (see \cite{CarNel,
CarNel1} for further details).  Our goal is to obtain three distinct
descriptions of this set of solutions---in terms of geometric structures,
Chern-Simons holonomies, and ADM moduli and momenta---and to understand
their relationships.  As we shall see in section 3, these three desriptions,
although classically equivalent, naturally lead to rather different
approaches to quantization.

The fundamental group of $\IR\!\times\!T^2$ has two generators,
$[\gamma_1]$ and $[\gamma_2]$, corresponding to the two independent
circumferences of the torus.  These satisfy the single relation
\beq
[\gamma_1]\cdot[\gamma_2] = [\gamma_2]\cdot[\gamma_1] .
\label{a32}
\eeq
The holonomy group $\cal M$ is therefore generated by two commuting
$\hbox{SO}(2,2)$ matrices, unique up to overall conjugation.  It is
somewhat more convenient to describe the holonomies as elements of the
covering group $\hbox{SL}(2,\IR)\!\times\!\hbox{SL}(2,\IR)$ \cite{NRZ};
I shall do so below.  Since the moduli space of the torus is
two-dimensional---it is parametrized by a single complex number
$\tau$---we expect from subsection \ref{ADM} that the phase space
should be four-dimensional; that is, we should find a four-parameter
family of holonomies $\rho$.

Let $\rho^\pm[\gamma_a]$ denote the two $\hbox{SL}(2,\IR)$ holonomies
corresponding to the curve $\gamma_a$.  An $\hbox{SL}(2,\IR)$ matrix
$S$ is called hyperbolic, elliptic, or parabolic according to
whether $|\Tr S|$ is greater than, equal to, or less than 2, and the
space of holonomies correspondingly splits into nine sectors.  It may
be shown that only the hyperbolic-hyperbolic sector corresponds to a
spacetime in which the $T^2$ slices are spacelike \cite{Ezawa,Ezawa2,LouMar}.
By suitable overall conjugation, the two generators of the holonomy
group can then be taken to be
\begin{eqnarray}
\rho^\pm[\gamma_1] = \left( \begin{array}{cc}
  e^{r_1^\pm/2} & 0 \\ 0 & e^{-r_1^\pm/2} \end{array}\right)
  \nonumber\\
\rho^\pm[\gamma_2] = \left( \begin{array}{cc}
  e^{r_2^\pm/2} & 0 \\ 0 & e^{-r_2^\pm/2} \end{array}\right) ,
\label{a33}
\end{eqnarray}
where the $r_a^\pm$ are four arbitrary parameters.

To obtain the corresponding geometry, we can use the quotient space
construction of subsection \ref{GS}.  Note first that three-dimensional
anti-de Sitter space is naturally isometric to the group manifold of
$\hbox{SL}(2,\IR)$.  Indeed, anti-de Sitter space can be represented as
the submanifold of flat $\IR^{2,2}$ (with coordinates $(X_1,X_2,T_1,T_2)$
and metric $dS^2 = dX_1^2 + dX_2^2 - dT_1^2 - dT_2^2$) on which
\beq
\hbox{det}|{\bf X}| = 1 ,
  \qquad {\bf X} = {1\over\ell}\left( \begin{array}{cc}
  X_1+T_1 & X_2+T_2\\ -X_2+T_2 & X_1-T_1 \end{array} \right) ,
\label{a34}
\eeq
i.e., ${\bf X}\in\hbox{SL}(2,\IR)$.  The quotient of anti-de Sitter space
by our holonomy group may be obtained by allowing the $\rho^+[\gamma_a]$
to act on $\bf X$ by left multiplication and the $\rho^-[\gamma_a]$ to
act by right multiplication.  It is not hard to show that the induced
metric on the resulting quotient space is
\begin{eqnarray}
ds^2 = dt^2 &-& {\ell^2\over4}
  \left[(r_1^+)^2 + (r_1^-)^2 +2r_1^+r_1^-\cos{2t\over\ell}\right] dx^2
  \nonumber\\ &\phantom{-}& \qquad
  - {\ell^2\over2}\left[r_1^+r_2^+ + r_1^-r_2^-
  + (r_1^+r_2^- + r_1^-r_2^+)\cos{2t\over\ell}\right] dxdy
  \\&\phantom{-}& \qquad\qquad - {\ell^2\over4}
  \left[(r_2^+)^2 + (r_2^-)^2 +2r_2^+r_2^-\cos{2t\over\ell}\right] dy^2 ,
  \nonumber
\label{a35}
\end{eqnarray}
where $x$ and $y$ are coordinates with period $1$.

It is easy to confirm that the metric \rref{a35} is indeed that of a space
of constant negative curvature.  To relate this result to the Chern-Simons
picture, observe that the corresponding triad and spin connection are
\begin{eqnarray}
e^0 &=&dt \nonumber\\
e^1&=& {\ell\over 2} \left[(r_1^+ - r_1^-)dx +(r_2^+ - r_2^-)dy\right]
 \sin{t\over{\ell}} \\
e^2&=& {\ell\over 2} \left[(r_1^+ + r_1^-)dx +(r_2^+ + r_2^-)dy\right]
 \cos{t\over{\ell}} \nonumber
\label{a36}
\end{eqnarray}
\begin{eqnarray}
\omega^{12}&=&0  \nonumber\\
\omega^{01}&=&-{1\over 2}\left[(r_1^+ - r_1^-)dx +(r_2^+ - r_2^-)dy\right]
 \cos{t\over{\ell}} \\
\omega^{02}&=&{1\over 2}\left[(r_1^+ + r_1^-)dx +(r_2^+ + r_2^-)dy\right]
 \sin{t\over{\ell}} . \nonumber
\label{a37}
\end{eqnarray}
These can be used to construct a pair of $\hbox{SL}(2,\IR)$ gauge fields
$A^{(\pm)a}$ as in \rref{a19}, and it is not hard to check that these have
vanishing field strength.  Conversely, the holonomies of the $A^{(\pm)a}$
may be shown to reproduce \rref{a33}, as required for consistency.

These expressions may in turn be related to the ADM formalism of subsection
\ref{ADM}.  For the metric \rref{a35}, the extrinsic curvature of a slice
of constant $t$ is
\beq
T = -{d\over dt}\ln\sqrt{g} = -{2\over\ell}\cot{2t\over\ell} ,
\label{a38}
\eeq
which is monotonic in $t$ in the range $(0,\pi\ell/2)$ and is
independent of $x$ and $y$.  Constant $t$ slices are thus also
slices of constant York time.  The modulus of a slice of constant $t$
is easily computed by comparing \rref{a35} to \rref{a26a}; one obtains
\beq
\tau = \left(r_1^-e^{it/\ell} + r_1^+e^{-{it/\ell}}\right)
 \left(r_2^-e^{it/\ell} + r_2^+e^{-{it/\ell}}\right)^{\lower2pt%
 \hbox{$\scriptstyle -1$}} .
\label{a39}
\eeq
The conjugate momentum $p = p^1 + ip^2$ can be similarly computed from
the extrinsic curvature of a constant $t$ slice, using \rref{a30}; it
takes the form
\beq
p= -{i\ell\over 2\sin{2t\over \ell}}\left(r_2^+e^{it/\ell}
 + r_2^-e^{-{it/\ell}}\right)^{\lower2pt%
 \hbox{$\scriptstyle 2$}} .
\label{a40}
\eeq
Finally, the ADM Hamiltonian $H$ may also be obtained from the metric,
using \rref{a29a}:
\beq
H = \left( T^2 + {4\over\ell^2}\right)^{-1} [\tau_2{}^2p\bar p]^{1/2}
  = {\ell^2\over4}\sin{2t\over\ell}(r_1^-r_2^+ - r_1^+r_2^-) .
\label{a41}
\eeq
In the limit of vanishing cosmological constant, these relations go over
to those of \cite{SCobserv} (see \cite{CarNel}).

Let us next consider the Poisson brackets among these variables.
{}From the brackets \rref{a22}, we find that
\beq
\{r_1^\pm,r_2^\pm\}=\mp {1\over\ell}, \qquad \{r^+_a,r^-_b\}=0 .
\label{a42}
\eeq
The corresponding brackets among the moduli and momenta $\tau$ and $p$
may be computed from \rref{a39} and \rref{a40}; we obtain
\beq
\{\tau,\bar p\} = \{\bar\tau, p\} = 2 ,\quad
\{\tau,p\} = \{\bar\tau,\bar p\} = 0 ,
\label{a43}
\eeq
in agreement with \rref{a30a}.  It may also be shown that these brackets
lead to a set of Hamilton's equations of motion that reproduce the
time dependence \rref{a39} of the moduli.  (See \cite{CarNel,Soda} for
a more detailed description of the dynamics.)  Our various descriptions
thus all agree, as they must.

It is also useful to exhibit the Poisson brackets among the traces of
the holonomies, which serve as a set of gauge-invariant observables.
Let
\begin{eqnarray}
R_1^\pm &=& {1\over2}\Tr \rho^\pm[\gamma_1] = \cosh{r_1^\pm\over2} ,
  \nonumber\\
R_2^\pm &=& {1\over2}\Tr \rho^\pm[\gamma_2] = \cosh{r_2^\pm\over2} ,\\
R_{12}^\pm &=& {1\over2}\Tr \rho^\pm[\gamma_1\cdot\gamma_2]
  = \cosh{(r_1^\pm+r_2^\pm)\over2} .\nonumber
\label{a44}
\end{eqnarray}
It is then not hard to check that
\beq
\{R_1^{\pm},R_2^{\pm}\}=\mp{1\over {4\ell}}(R_{12}^{\pm}-
 R_1^{\pm}R_2^{\pm}) \quad \hbox{\it and cyclical permutations},
\label{a45}
\eeq
reproducing the Poisson algebra of Nelson, Regge, and Zertuche \cite{NRZ}.

Finally, let us consider the action of the torus mapping class group.
This group is generated by two Dehn twists, which act on $\pi_1(T^2)$
by
\begin{eqnarray}
&S&:\gamma_1\rightarrow \gamma_2^{-1},\hphantom{\gamma_2}
    \qquad \gamma_2\rightarrow\gamma_1\nonumber\\
&T&:\gamma_1\rightarrow\gamma_1\cdot\gamma_2 ,
    \qquad \gamma_2\rightarrow\gamma_2 .
\label{a46}
\end{eqnarray}
These transformations act on the parameters $r_a^\pm$ as
\begin{eqnarray}
&S&:r_1^{\pm}\rightarrow r_2^{\pm},\hphantom{+ r_2^\pm r}\!\qquad
    r_2^{\pm}\rightarrow - r_1^{\pm}\nonumber\\
&T&:r_1^{\pm}\rightarrow r_1^{\pm} + r_2^{\pm},\qquad
    r_2^{\pm}\rightarrow r_2^{\pm} ,
\label{a47}
\end{eqnarray}
and on the ADM moduli and momenta as
\begin{eqnarray}
&S&: \tau\rightarrow -{1\over\tau} ,\hphantom{1}\qquad
     p\rightarrow \bar \tau^2 p \nonumber\\
&T&: \tau\rightarrow \tau+1 ,\qquad p\rightarrow p .
\label{a48}
\end{eqnarray}
It is not hard to show that these transformations are consistent with
the relationships between the ADM and holonomy variables, and that they
preserve all Poisson brackets.

\subsection{The (2+1)-Dimensional Black Hole \label{BH}}

I will finish this section by describing another exact solution
with a negative cosmological constant, the (2+1)-dimensional black
hole.  The discovery of this solution by Ba\~nados, Teitelboim, and
Zanelli (BTZ) \cite{BTZ,BHTZ} came as a surprise, since it had been
generally assumed that physically realistic solutions required
the full dynamics of 3+1 dimensions.  Indeed, the (2+1)-dimensional
black hole differs from its (3+1)-dimensional counterpart in one
important way: since the spacetime curvature is constant in 2+1
dimensions, there can be no curvature singularity at the origin.
Nevertheless, the BTZ solution shares many of the essential
features of a realistic black hole, including an event horizon, an
inner horizon (in the rotating case), and thermodynamic properties.

Like the torus universe, the black hole has a geometry that can be
described in several different languages.  The relationship to the
ordinary Schwarzschild and Kerr black holes is most easily exhibited
in the metric formalism.  The BTZ metric is
\beq
ds^2 = -( N^\perp)^2dt^2 + f^{-2}dr^2
  + r^2\left( d\phi + N^\phi dt\right)^2
\label{a49}
\eeq
with lapse and shift functions
\beq
N^\perp = f
  = \left( -M + {r^2\over\ell^2} + {J^2\over4r^2} \right)^{1/2} ,
  \qquad N^\phi = - {J\over2r^2} \ .
\label{a50}
\eeq
Here, the time coordinate $t$ is not the York time of subsection \ref{ADM},
but is rather the ``Killing time,'' the displacement along a timelike
Killing vector at spatial infinity.  (The analysis in terms of York time
is possible, but considerably more complicated \cite{CarHos}.)  It is
straightforward to check that the metric \rref{a49} satisfies the Einstein
field equations with a cosmological constant $\Lambda=-1/\ell^2$.

When  $M\!>\!0$ and $|J|\!\leq\!M\ell$, this solution has an outer event
horizon at $r=r_+$ and an inner horizon at $r=r_-$, where
\beq
r_\pm^2={M\ell^2\over 2}\left \{ 1 \pm
\left [ 1 - \left({J\over M\ell}\right )^2\right ]^{1/2}\right \} ,
\label{a51}
\eeq
i.e.,
\beq
M={r_+^2+r_-^2\over\ell^2}, \quad J={2r_+ r_-\over\ell} \ .
\eeq
The parameters $M$ and $J$ can be studied either by looking at spatial
integrals at infinity \cite{BHTZ} or by examining quasilocal expressions
at a finite spatial boundary \cite{BrownMann}; either analysis shows
that they are simply the mass and angular momentum of the black hole.

Since the black hole metric has constant negative curvature, it must be
at least locally isometric to anti-de Sitter space.  For the region
$r\!>\!r_+$, this isometry may be exhibited by means of a coordinate change
(the corresponding transformations for $r\!<\!r_+$ are given in \cite{BHTZ};
see also \cite{CarTeit}):
\begin{eqnarray}
x &=& \left({r^2-r_+^2\over r^2-r_-^2}\right)^{1/2}
      \cosh\left( {r_+\over\ell^2}t - {r_-\over\ell}\phi \right)
      \exp\left\{ {r_+\over\ell}\phi - {r_-\over\ell^2}t \right\}
      \nonumber \\
y &=& \left({r^2-r_+^2\over r^2-r_-^2}\right)^{1/2}
      \sinh\left( {r_+\over\ell^2}t - {r_-\over\ell}\phi \right)
      \exp\left\{ {r_+\over\ell}\phi - {r_-\over\ell^2}t \right\} \\
z &=& \left({r_+^2-r_-^2\over r^2-r_-^2}\right)^{1/2}
      \exp\left\{ {r_+\over\ell}\phi - {r_-\over\ell^2}t \right\} ,
      \nonumber
\label{a52}
\end{eqnarray}
for which the metric becomes
\beq
ds^2 = {\ell^2\over z^2} (dx^2 - dy^2 + dz^2) .
\label{a53}
\eeq
This expression may be recognized as the standard Poincar\'e metric
for anti-de Sitter space.  Note, however, that periodicity in the
Schwarzschild angular coordinate $\phi$ requires that we identify points
under the action $\phi\!\mapsto\!\phi+2\pi$, that is,
\begin{eqnarray}
&(&x,y,z)\sim \\
&\phantom{.}&\left(
e^{2\pi r_+/\ell}(x\cosh{2\pi r_-\over\ell} - y\sinh{2\pi r_-\over\ell}),\,
e^{2\pi r_+/\ell}(y\cosh{2\pi r_-\over\ell} - x\sinh{2\pi r_-\over\ell}),\,
e^{2\pi r_+/\ell}z\right) \nonumber
\label{a54}
\end{eqnarray}
These identifications are an isometry of the metric \rref{a53}, and
thus represent an element of the isometry group $\hbox{SL}(2,\IR)\!\times\!
\hbox{SL}(2,\IR)$ of anti-de Sitter space.  This element is, in fact,
the holonomy of the geometric structure of the black hole in the sense
of subsection \ref{GS}.  The corresponding $\hbox{SL}(2,\IR)$ matrices
$\rho^\pm$ may be obtained from the group action described after
equation \rref{a34}; a bit of computation gives
\beq
\rho^\pm = \left( \begin{array}{cc}
 e^{\pi r_\pm/\ell} & 0 \\ 0 & e^{-\pi r_\pm/\ell}
 \end{array}\right) .
\label{a55}
\eeq

The $\hbox{SL}(2,\IR)$ Chern-Simons connections \rref{a19} can also be
obtained from \rref{a53}--\rref{a54}, and it is a simple exercise to check
that the holonomy of the connection is the same as the holonomy of the
geometric structure.  This holonomy was originally computed in the
Schwarzschild coordinates \rref{a49} in reference \cite{Manna}; the
resulting expression differs from \rref{a55} by a complicated overall
conjugation.  Note that the Poisson brackets for the black hole spacetime
are rather mysterious: the parameters $r_+$ and $r_-$ are independent,
and seem to have no canonical conjugates.  We shall return to this issue
in subsection \ref{Therm}.

\section{Quantum Gravity in 2+1 Dimensions \label{Quant}}
\setcounter{footnote}{0}

The main goal of studying general relativity in (2+1) dimensions is to
gain insight into the problems of quantum gravity.  It may therefore seem
that I have spent an inordinate amount of time on the details of the
classical theory.  We shall see, however, that the three classical
descriptions of the last section lead very directly to three approaches
to quantization.  Indeed, one of the main lessons of (2+1)-dimensional
gravity seems to be that a thorough understanding of the classical
solutions is crucial for the formulation of a quantum theory.

Before starting in on the problem of quantization, it is worth recalling
why quantum gravity is so hard.   The difficulties are partly technical:
general relativity is a complicated, nonlinear theory, and approximation
methods that work elsewhere simply break down.  In particular, general
relativity is perturbatively nonrenormalizable, and while we know a few
examples of nonrenormalizable theories that can be sensibly quantized,
the general problem is poorly understood.

Beyond these technical failures, however, lie the basic conceptual
problems that plague quantum gravity.  Conventional quantum theory starts
with a fixed, passive spacetime background that provides a setting in
which particles and fields interact.  According to general relativity,
however, spacetime is itself dynamical, and much of the conventional
framework becomes, at best, ambiguous.  Without a fixed definition of
time, we do not know how to describe dynamics or interpret probabilities.
Without an {\em a priori\/} distinction between past, present, and future,
we do not how to impose causality.  Fundamentally, we do not understand
what it means to quantize the structure of spacetime.

The usefulness of (2+1)-dimensional gravity comes from the fact that
it eliminates the technical problems while preserving the conceptual
foundations.  We have seen that for typical topologies, (2+1)-dimensional
general relativity has only finitely many degrees of freedom.  Quantum
field theory is thus reduced to quantum mechanics, and the problem of
nonrenormalizability disappears.  On the other hand, (2+1)-dimensional
gravity is still a diffeomorphism-invariant theory of spacetime geometry,
and most of the basic conceptual issues of the full theory remain
unchanged.

To describe the quantization of (2+1)-dimensional general relativity, I
will work backwards through the classical descriptions of the previous
section, starting with the ADM formalism and ending with the quantum
mechanics of geometric structures.  A final subsection will briefly
address some other approaches to quantization, including path integral
methods and lattice approaches.

\subsection{Reduced Phase Space Quantization \label{RPS}}

Perhaps the simplest approach to quantum gravity in 2+1 dimensions
\cite{SCobserv,HosNak2} starts from the reduced phase space action
\rref{a29}, which was obtained by solving the constraints in the
metric formalism.  This action describes a finite-dimensional system
in  classical mechanics, albeit one with a complicated, time-dependent
Hamiltonian.  We know, at least in principle, how to quantize such a
system: we simply replace the Poisson brackets \rref{a30a} with
commutators,
\beq
[ \hat m_\alpha, \hat p^\beta ] = i\hbar\delta^\beta_\alpha ,
\label{b1}
\eeq
represent the momenta as derivatives,
\beq
p^\alpha = {\hbar\over i}{\partial\ \over\partial m_\alpha} ,
\label{b2}
\eeq
and choose our wave functions to be square integrable functions
$\psi(m_\alpha,T)$ that evolve according to the Schr\"odinger equation
\beq
i\hbar{\partial\psi(m_\alpha,T)\over\partial T} = \hat H\psi(m_\alpha,T) ,
\label{b3}
\eeq
where the Hamiltonian $\hat H$ is obtained from \rref{a29a} by some
suitable operator ordering.  Invariance under the mapping class group
can be incorporated by demanding that the $\psi(m_\alpha,T)$ transform
under a suitable representation of ${\cal D}(M)$; a similar requirement
can help determine the operator ordering in the Hamiltonian operator,
although some ambiguities will remain.

For spatial surfaces of genus $g\!>\!2$, the complexity of the constraint
\rref{a28}---which must be solved to determine $\hat H$---may make this
approach to quantization impractical.  A perturbative expression for
$\hat H$ might still exist, however, and Puzio has suggested that the
Gauss map could provide a useful tool \cite{Puzio}.  For spaces of genus
2, there is some hope that the representation of $\Sigma$ as a
hyperelliptic surface could lead to an exact solution of the constraint;
and for the genus 1 case, the exact expression for $\hat H$ is already
known.

Indeed, the classical Hamiltonian for the torus universe is given by
equation \rref{a41}, which yields, up to operator ordering ambiguities,
\beq
\hat H = {\hbar\over\sqrt{T^2-4\Lambda}}\Delta_0^{1/2} ,
\label{b4}
\eeq
where
\beq
\Delta_0 = -\tau_2^{\ 2}\left( {\partial^2\ \over\partial \tau_1^{\ 2}} +
  {\partial^2\ \over\partial \tau_2^{\ 2}}\right)
\label{b5}
\eeq
is the ordinary scalar Laplacian for the constant negative curvature
Poincar\'e metric on moduli space.  This  Laplacian is invariant under
the modular transformations \rref{a48}, and its invariant eigenfunctions,
known as weight zero Maass forms, have been studied rather carefully
in the mathematical literature \cite{Maass}.  The behavior of the
corresponding wave functions is discussed by Puzio \cite{Puzio2}, who
argues that they are well-behaved at the ``edges'' of moduli space.

While the choice \rref{b5} of operator ordering is not unique, the number
of possible alternatives is smaller than one might naively expect.  The
key restriction is diffeomorphism invariance: the eigenfunctions of the
Hamiltonian should transform under a one-dimensional unitary representation
of the mapping class group.  The representation theory of the modular group
\rref{a48} has been studied extensively \cite{Maass2,Maass3}; one finds
that the possible inequivalent Hamiltonians are all of the form \rref{b4},
but with $\Delta_0$ replaced by\footnote{It is argued in \cite{dirac} that
the natural choice of ordering in Chern-Simons quantization corresponds to
$n=1/2$.}
\beq
\Delta_n = -\tau_2^{\ 2}\left( {\partial^2\ \over\partial \tau_1^{\ 2}} +
  {\partial^2\ \over\partial \tau_2^{\ 2}}\right)
  + 2in\tau_2{\partial\ \over\partial \tau_1} + n(n+1) , \quad 2n\in\IZ ,
\label{b6}
\eeq
the Maass Laplacian acting on automorphic forms of weight $n$.  (See
\cite{ordering} for details of the required operator orderings.)  Note
that when written in terms of the momentum $p$, the operators $\Delta_n$
differ from each other by terms of order $\hbar$, as expected for operator
ordering ambiguities.  Nevertheless, the various choices of ordering may
have drastic effects on the physics, since the spectra of the various Maass
Laplacians are quite different.

This ordering ambiguity may be viewed as arising from the structure of
the classical phase space.  The torus moduli space is not a manifold, but
rather has orbifold singularities, and quantization on an orbifold
is generally not unique.  Since the space of solutions of the Einstein
equations in 3+1 dimensions has a similar orbifold structure \cite{Isen},
we might expect a similar ambiguity in realistic (3+1)-dimensional
quantum gravity.

The quantization presented here is an example of what Kucha{\v r} calls
an ``internal Schr\"odinger interpretation'' \cite{Kuchar}.  It appears
to be a completely self-consistent approach, and like ordinary
quantum mechanics, it automatically has the correct classical limit
on the reduced phase space of subsection \ref{ADM}.  The key drawback of
this approach is that it relies on a {\em classical\/} choice of time
slicing.  The analysis of subsection \ref{ADM} required that we choose
the York time slicing from the start, and it is not at all clear that a
different choice would lead to an equivalent quantum  theory.  In
other words, it is not clear that this approach to quantum gravity
preserves general covariance.

The problem may be rephrased as a statement about the kinds of questions
we can ask in this quantum theory.  The model naturally allows us to
compute the transition amplitude between the spatial geometry of a time
slice of constant mean curvature $-\Tr K\!=\!T_1$ and the geometry of a later
slice of constant mean curvature $-\Tr K\!=\!T_2$.  But it is not clear how
to ask for transition amplitudes among other spatial slices, on which
$\Tr K$ is not constant; such questions would require a different classical
time slicing, and hence a different---and perhaps inequivalent---quantum
theory.

To try to escape this difficulty, we next turn to an alternative
approach to quantization, which starts from the Chern-Simons formalism.

\subsection{Chern-Simons Quantum Theory \label{CSQT}}

In the first-order formulation of subsection \ref{CS}, the fundamental
gauge-invariant observables of quantum gravity are the traces
$\Tr\rho[\gamma]$ of the holonomies.  These provide an overcomplete
set of coordinates for the space of classical solutions, and it is
not obvious that the entire set of Poisson brackets can be made into
commutators of operators.  This problem has been studied systematically
by Nelson and Regge \cite{NR1,NR3,NR2,NRZ,NR4,NR5,NR6}, who demonstrate
the existence of a well-behaved subalgebra of traces for which commutators
can be consistently defined.

For the case of the torus universe, this approach is fairly straightforward.
To quantize the algebra \rref{a45}, we proceed as follows:
\setlength{\leftmargini}{1.2em}
\begin{enumerate}
\item We replace the classical Poisson brackets $\{\,,\,\}$ by commutators
$[\,,\,]$, with the rule
$$
[x,y]= xy-yx =i \hbar \{x,y\} ;
$$
\item On the right hand side of \rref{a45}, we replace the product by
the symmetrized product,
$$
xy \to {1\over2} (xy +yx) .
$$
\end{enumerate}
The resulting algebra is defined by the relations
\beq
\hat R_1^{\pm}\hat R_2^{\pm}e^{\pm i \theta}
  - \hat R_2^{\pm}\hat R_1^{\pm} e^{\mp i \theta}=
  \pm 2i\sin\theta\, \hat R_{12}^{\pm} \quad \hbox{\it and cyclical
  permutations}
\label{b7}
\eeq
with
\beq
\tan\theta= -{\hbar\over8\ell} \ .
\label{b8}
\eeq
The algebra \rref{b7} is not a Lie algebra, but it is related to the
Lie algebra of the quantum group $\hbox{SU}_q(2)$ \cite{NRZ}, with
$q=\exp{4i\theta}$.

In terms of the parameters $r_a^\pm$ of subsection \ref{Tor}, this
algebra can be represented by \cite{CarNel,CarNel1}
\beq
\hat R_1^{\pm} = \sec\theta\cosh{\hat r_1^{\pm}\over 2} ,\quad
\hat R_2^{\pm} = \sec\theta\cosh{\hat r_2^{\pm}\over 2} ,\quad
\hat R_{12}^{\pm} = \sec\theta\cosh{(\hat r_1^{\pm}+\hat r_2^{\pm})\over 2}
\label{b9}
\eeq
with
\beq
[\hat r_1^{\pm}, \hat r_2^{\pm}] = \pm 8i\theta,
  \qquad[\hat r^+_a,\hat r^-_b] = 0 .
\label{b10}
\eeq
For $\Lambda$ small, these commutators differ from the naive quantization
of the classical brackets \rref{a42},
\beq
[\hat r_1^\pm, \hat r_2^\pm] = \mp {i\hbar\over\ell} ,
\label{b11}
\eeq
by terms of order $\hbar^3$.

We must next implement the action of the modular group \rref{a47}
on the operators $\hat R^\pm_a$.  The action that preserves the
algebraic relations \rref{b7} is
\begin{eqnarray}
&S&:\hat R_1^{\pm}\rightarrow \hat R_2^{\pm},
  \quad \hat R_2^{\pm}\rightarrow \hat R_1^{\pm},
  \quad \hat R_{12}^{\pm}\rightarrow
        \hat R_1^{\pm} \hat R_2^{\pm} + \hat R_2^{\pm} \hat R_1^{\pm}
        - \hat R_{12}^{\pm}\nonumber\\
&T&:\hat R_1^{\pm}\rightarrow \hat R_{12}^{\pm},\quad
  \hat R_2^{\pm}\rightarrow \hat R_2^{\pm}, \quad
  \hat R_{12}^{\pm}\rightarrow \hat R_{12}^{\pm}\hat  R_2^{\pm}
        + \hat R_2^{\pm} \hat R_{12}^{\pm} - \hat R_1^{\pm} ,
\label{b12}
\end{eqnarray}
which may be recognized as a particular factor ordering of the classical
group action.

This approach to quantization is an example of what Kucha{\v r} calls
``quantum gravity without time'' \cite{Kuchar}.  Like the quantization
of subsection \ref{RPS}, it appears to be self-consistent, but like
that approach, it suffers from an important deficiency.  In this case,
the observables $\hat R_a$ characterize the entire spacetime at once,
and are therefore time-independent constants of motion.  But the classical
solutions to (2+1)-dimensional gravity, even for the simple $\IR\!\times
\!T^2$ topology, are most certainly {\em not\/} static.  Our operators
have somehow lost track of the dynamics of spacetime.

As in subsection \ref{RPS}, the problem can be phrased as a limitation
on the kinds of questions we can ask.  In this quantization, we can
ask about the overall spacetime geometry, but we have apparently lost
the ability to describe dynamics.  This ``frozen time'' problem is
characteristic of generally covariant theories: translations in
(coordinate) time are diffeomorphisms, and cannot be seen by looking
at diffeomorphism-invariant observables.  Rovelli has suggested a solution
to this difficulty, based on the idea of ``evolving constants of motion''
\cite{Rovelli1,Rovelli2}.  To see this method in action, it is useful to
turn to yet another approach to quantum gravity, that of covariant
canonical quantization.

\subsection{Covariant Canonical Quantization \label{CCQ}}

The starting point of covariant canonical quantization \cite{AshMag,Crn,
AshBom,Wald} is the simple but profound observation that the phase space
of any well-behaved classical theory is isomorphic to the space of
classical solutions.  Specifically, let $\cal C$ be an arbitrary (but
fixed) Cauchy surface.  Then a point in the phase space determines
initial data on $\cal C$, which in turn determine a unique solution,
while, conversely, a classical solution restricted to $\cal C$ determines
a point in the phase space.  This observation suggests a solution of the
old ``covariant vs.\ canonical'' debate in quantum gravity \cite{Isham},
by offering a manifestly covariant approach to canonical quantization.

For (2+1)-dimensional gravity on a spacetime with the topology $\IR\!
\times\!\Sigma$, we saw in subsection \ref{GS} that the space of
classical solutions of the vacuum field equations is the simply the
space $T^*{\cal N}(\Sigma)/{\cal D}(\Sigma)$ of holonomies of geometric
structures.  As a cotangent bundle, this space has a natural
symplectic structure, and it follows from the results of subsection
\ref{CS} (see equations \rref{a13}--\rref{a14} and \rref{a22}) that
this structure is closely related to the symplectic structure of the
classical Poisson brackets.  When $\Lambda\!\ne\!0$, the cotangent
bundle structure disappears, but the structure of the fundamental
group of a surface induces a symplectic structure on ${\cal M}/{\cal D}$,
which is again closely related to the classical Poisson bracket structure.
As Goldman has shown \cite{Goldman4,GoldHam}, this symplectic structure
may be described abstractly in terms of Poisson brackets among
intersecting loops on $\Sigma$, with added relations that depend on
the ``gauge group'' $G$ of the geometric structure.  For any particular
parametrization of $\cal M$, explicit expressions for the brackets
among parameters may be determined from Chern-Simons quantization or
from the operator algebras of Nelson and Regge. For the $\IR\!\times\!T^2$
universe, for example, the $r_a^\pm$ can be interpreted as coordinates
on $\cal M$, and the brackets \rref{a42} determine a symplectic structure
on this space.

To quantize this covariant phase space, we must rewrite a suitable
subalgebra of classical Poisson brackets as commutators.  We are thus
led to a quantum theory in which wave functions are functions of the
holonomies $\hat\rho\!\in\!{\cal N}$, or, if $\Lambda\!\ne\!0$, of
a suitably chosen set of half of the coordinates of $\cal M$.  For
the torus universe, for instance, we obtain operators $\hat r_a^\pm$
satisfying the commutation relations \rref{b11},\footnote{We could
replace these commutators with those of \rref{b10} by ``rescaling
Planck's constant.''} and wave functions of the form $\psi(r_1^+,r_1^-)$.
(To recover the results of reference \cite{SCobserv} in the limit
$\Lambda\!\rightarrow\!0$, we should choose a different ``polarization,''
writing wave functions as functions of the commuting variables $u_1$
and $u_2$, where $u_a = (r^+_a-r^-_a)/2$; see \cite{CarNel} for details.)

Like the Chern-Simons quantization of the last subsection, covariant
canonical quantization leads to a ``frozen time'' formalism, in which
the fundamental operators are constants of motion that describe an
entire classical spacetime.  It should be clear, however, that this is
not just a problem of the quantization: our {\em classical\/} description
of the phase space is already time-independent.  But we know how to solve
this problem in the classical theory: we simply make use of the isomorphism
between the covariant canonical description and the conventional
time-dependent description of the phase space.

Let me rephrase this idea, as it applies to (2+1)-dimensional gravity, in
a slightly less abstract manner.  Given a (time-independent) holonomy
$\rho$ of a geometric structure, we can reconstruct a unique classical
spacetime.  Within that spacetime, we can ask ordinary dynamical
questions: for instance, what are the moduli of the slice of York
time $-\Tr K = T$?  Indeed, for the torus universe, this is precisely
the content of equation \rref{a39}, which gives the modulus of a torus
of fixed $T$ in terms of the coordinates $r_a^\pm$ of the geometric
structure.  Similarly, given any other time slicing, the question
``What are the moduli of the time $T'$ slice?'' has a unique classical
answer, depending only on $\rho$ and $T'$.

To carry this construction over to the quantum theory, we need merely
convert equation \rref{a39}, and its counterparts for other
time-dependent observables, into operator relations.  For example,
we can define a one-parameter family of operators $\hat\tau(t)$ by
\beq
\hat\tau(t) = \left(\hat r_1^-e^{it/\ell} + \hat r_1^+e^{-{it/\ell}}\right)
 \left(\hat r_2^-e^{it/\ell} + \hat r_2^+e^{-{it/\ell}}\right)^{\lower2pt%
 \hbox{$\scriptstyle -1$}} ,
\label{b13}
\eeq
where $t$ is now simply a label whose physical significance comes from
the classical limit.  Similarly, we can use equation \rref{a40} to
define a second one-parameter family of operators that represent the
ADM momenta on a slice of constant York time,
\beq
\hat p(t) = -{i\ell\over 2\sin{2t\over \ell}}\left(\hat r_2^+e^{it/\ell}
 + \hat r_2^-e^{-{it/\ell}}\right)^{\lower2pt%
 \hbox{$\scriptstyle 2$}} ;
\label{b14}
\eeq
and from \rref{a41}, we can define a third family representing the
Hamiltonian that generates translations in York time,
\beq
\hat H(t) = {\ell^2\over4}\sin{2t\over\ell}
  (\hat r_1^-\hat r_2^+ - \hat r_1^+\hat r_2^-) .
\label{b15}
\eeq
It may be checked that these operators obey the standard Heisenberg
equations of motion, and that they transform correctly under the
mapping class group.  (The requirement of simple behavior under the
action of the mapping class group was, in fact, a key ingredient in
determining the operator orderings.)

If we are given a different classical time slicing, labeled by a new
coordinate $T'$, we can repeat this procedure, using the classical
solutions to construct new operators $\{\hat\tau'(T'),\hat p'(T'),
\hat H'(T')\}$, which again act on the Hilbert space of the covariant
canonical quantum theory.  From this viewpoint, the ``frozen time''
problem is simply a sign that covariant canonical quantization leads
to a Heisenberg picture, in which wave functions {\em are\/}
time-independent; the dynamics now lies in our choice of families
of operators.  Of course, this construction involves operator
ordering ambiguities, and exact expressions for the ``time-dependent''
operators require our knowing the exact solutions of the classical
field equations.  But these problems are in some sense technical, and
perhaps not fundamental.

Finally, let me relate the results of this subsection to those of
subsections \ref{RPS} and \ref{CSQT}.  The connection to Chern-Simons
quantization is straightforward: equation \rref{b9} expresses the
Chern-Simons operators in terms of the $\hat r_a^\pm$, and if we choose
the commutators \rref{b10}, we recover the Nelson-Regge operator
algebra.  To connect covariant canonical quantization to the ``internal
Schr\"odinger interpretation'' of subsection \ref{RPS}, we must transform
from a Heisenberg picture to a Schr\"odinger picture by diagonalizing
the operators $\hat\tau(t)$.  It is easy to construct a formal integral
transform to do so (see \cite{dirac} and \cite{Ezawa3} for details); one
obtains wave functions that formally obey the Schr\"odinger equation
\rref{b3}, but with a Hamiltonian determined by \rref{b6} with
$n=1/2$.\footnote{Other operator orderings in \rref{b13} lead to
other values of $n$; see \cite{ordering}.}  Peldan has recently
pointed out one potential problem with this transformation, however
\cite{Peldan}: the torus mapping class group does not act properly
discontinuously on the ``configuration spaces'' parametrized by $(r_1^+,
r_1^-)$ or $(u_1,u_2)$, so invariant wave functions cannot be taken to
be smooth functions of these configuration space variables.  I do not
know a resolution to this problem; it may be that the requirement of
smoothness is too strong, but further work is needed.

\subsection{Other Approaches to Quantization \label{Other}}

Let me complete this section by briefly mentioning a few other approaches
to quantum gravity in 2+1 dimensions.  I cannot hope to do these methods
justice in the limited space available; this subsection should be
interpreted as a ``reading list.''  For an earlier review of some of
these quantum theories, see reference \cite{Six}.

\begin{enumerate}\addtolength{\itemsep}{2ex}\addtolength{\leftmargin}{-1em}
\item {\bf The Wheeler-DeWitt Equation}\\

One conventional approach to quantum gravity is that of the Wheeler-DeWitt
equation.  This approach starts with the ADM decomposition of subsection
\ref{ADM}, but rather than choosing a time slicing and solving the
Hamiltonian constraint ${\cal H}\!=\!0$ classically, we impose it as a
Klein-Gordon-like equation of motion restricting physical states $\Psi
[\bar g_{ij},\lambda]$.  For the torus universe, this constraint takes
the form of a functional differental equation \cite{WdW}
\begin{eqnarray}
\Biggl\{
&{\displaystyle{1\over8}}&{\delta\ \over\delta\lambda}
  e^{-2\lambda}{\delta\ \over\delta\lambda}
  + {1\over2}e^{-2\lambda}\Delta_0 + 2 \bar\Delta\lambda
  - 2e^{-2\lambda}Y^i[\pi]\bar\Delta Y_i[\pi] \\
  &+& 2e^{-2\lambda}\delbar_i\Biggl[\left(2p^{ij}
      + \delbar^iY^j[\pi] + \delbar^jY^i[\pi]
      - \bar g^{ij}\delbar_kY^k[\pi]\right)Y_j[\pi]\Biggr]
  \Biggr\}\Psi[\lambda,\tau]
  = 0 ,\nonumber
\label{b16}
\end{eqnarray}
where $Y_i[\pi]$ is given by
\beq
Y_i = -{1\over2} \bar\Delta^{-1}
  \left[{1\over\sqrt{\bar g}}e^{2\lambda}
  \delbar_i\left(e^{-2\lambda}\pi\right)\right]
\label{b17}
\eeq
and
\beq
\pi = -{i\over2}{\delta\ \over\delta\lambda}, \quad
p^{ij} = - {i\over\sqrt{\bar g}}{\delta\ \over\delta\bar g_{ij}} \ .
\label{b17a}
\eeq
Note the appearance of nonlocal terms, those involving the $Y_i$, which
arise from solving the momentum constraints ${\cal H}_i=0$.  The existence
of such nonlocal behavior was first pointed out by Henneaux \cite{Henn2}.
The properties of equation \rref{b16}, and proposals for an inner product
for the $\Psi[\lambda,\tau]$, are discussed in \cite{WdW}; the results
do not appear to be equivalent to those of any of the quantizations
discussed above, but operator ordering ambiguities in \rref{b16} are
severe enough that we cannot be certain of this claim.  See also
\cite{Visser,Banks} for ``gauge-fixed'' versions of the Wheeler-DeWitt
equation, which are essentially minisuperspace models that eliminate
the nonlocal terms in \rref{b16}.

\item {\bf Loop Variables}\\

The Chern-Simons variables of subsection \ref{CS} are, roughly speaking,
the (2+1)-dimensional analogues of Ashtekar's variables in 3+1 dimensions
\cite{Ashbook,Ash2}.  Work on these variables has led to an interesting
``dual'' representation, the loop representation \cite{RovSmol}, in
which wave functions and operators are functionals of loops.  Wave
functions in the loop representation are related to those in the
``connection representation'' by the loop transform, defined formally
as the functional integral
\beq
\tilde\Psi[\gamma] = \int[d\omega]\,\hat\rho[\gamma][\omega]\Psi[\omega] ,
\label{b18}
\eeq
where $\hat\rho[\gamma][\omega]$ is the Wilson loop \rref{a12a}.  In 2+1
dimensions (with vanishing cosmological constant), $\omega$ is a flat
connection, and the wave function $\tilde\Psi[\gamma]$ depends only on
the homotopy class of $\gamma$.  For the torus, in particular, any
loop is determined up to homotopy by two winding numbers---in terms of
the basis of subsection \ref{Tor},
\beq
\gamma\approx\gamma_1^{n_1}\cdot\gamma_2^{n_2} .
\label{b19}
\eeq
Wave functions in the loop representation are thus functions $\psi(n_1,
n_2)$ of a pair of integers, and the loop transform reduces to an
ordinary Laplace transform \cite{Marolf,Smolin}.  As Marolf has shown
\cite{Marolf}, this transform is {\em not\/} an isomorphism, and the
loop representation for $\IR\!\times\!T^2$ is consequently inequivalent
to the ``connection representation'' of Chern-Simons quantization.
On the other hand, the integration measure in \rref{b18} can be
redefined---although in a rather unpleasant way---to make the transform
an isomorphism \cite{AshLoll}.  There are thus (at least) two loop
representations, one equivalent to more conventional quantization
and one not.

\item {\bf Lattice Methods}\\

We saw in section 2 that the solutions of the vacuum Einstein field
equations in 2+1 dimensions are flat.  As a consequence, lattice techniques,
in which spacetime is approximated by flat polyhedra, are in some sense
exact.  A number of authors, including Waelbroeck \cite{Wael,waela,waelb},
't~Hooft \cite{tH2,tH3}, and Newbury and Unruh \cite{NewUn}, have taken
advantage of this fact to construct simple lattice models of classical
and quantum gravity based on polyhedral decompositions of space or
spacetime.  These models are for the most part based on the first-order
formalism, and should yield quantum theories equivalent to those of
subsection \ref{CS}, but the relationships have not been fully explored.

A second set of lattice models come from Regge calculus.  These are
based on the old observation of Ponzano and Regge \cite{PonReg,Hass} that
the (Euclidean) three-dimensional gravitational path integral in Regge
calculus variables can be rewritten as a particular sum over 6-j symbols.
Turaev and Viro succeeded in regularizing this sum by replacing ordinary
6-j symbols with those of a quantum group \cite{TurVir}, and a number
of authors \cite{Oog1,Oog2,ArchWill,Rovlat,Iwa,Fox} have related the
states, observables, and expectation values of this model to those of
other approaches to quantization.  As discussed in reference \cite{Six},
there are problems with such comparisons, arising from the fact that
the Ponzano-Regge action is based on {\em Euclidean\/} gravity; the
geometric structures of a manifold with a positive definite metric are
quite different from those of a manifold with a Lorentzian metric.
Recently, however, Barrett and Foxon have proposed a new approach to
the Ponzano-Regge model that may be equivalent to (2+1)-dimensional
general relativity with Lorentzian signature \cite{Barrett}; further
work on this model could be very interesting.

\item {\bf Path Integrals}\\

I will finish this section with a brief discussion of the path integral
approach to (2+1)-dimensional gravity, as first developed by Witten
\cite{Witb}.  Our starting point is the first-order action \rref{a8}.
$I_{\hbox{\scriptsize grav}}$ is not a free field action---it is not
quadratic in the fields---but it is nevertheless possible to evaluate
the path integral exactly.  Roughly speaking, the integral over the triad
$e$ gives a delta functional $\delta[R[\omega]]$, where $R^a[\omega]$ is
the curvature defined by equation \rref{a12}; the delta functional then
permits an exact evaluation of the remaining integral over $\omega$.  When
the Faddeev-Popov determinants that arise from gauge-fixing the ISO(2,1)
transformations \rref{a9}--\rref{a10} are taken into account, one obtains
a partition function consisting of a combination of determinants known as
the Ray-Singer torsion,
\beq
T[\bar\omega] =
{(\hbox{det}'\bar\Delta_{(3)})^{3/2}
(\hbox{det}'\bar\Delta_{(1)})^{1/2}\over
(\hbox{det}'\bar\Delta_{(2)})} .
\label{b20}
\eeq
Here $\bar\omega$ is a flat connection---the zero of the argument of the
delta functional---and $\bar\Delta_{(k)}$ is the Laplacian $*D_{\bar\omega}
\!*\!D_{\bar\omega} + D_{\bar\omega}\!*\!D_{\bar\omega}*$ acting on
$k$-forms, where $D_{\bar\omega}$ is the covariant derivative.  This
combination of determinants is a topological invariant of a flat bundle
over the spacetime manifold $M$ \cite{RaySinger}, and is proportional to
the Reidemeister torsion, a combinatorial invariant of flat bundles
\cite{Milnor,Luck}.  This same combination of determinants can also be
obtained by means of a non-Abelian Hodge decomposition of the ``gauge''
fields $e$ and $\omega$ \cite{GegKun}, or by a BRST analysis of the
path integral \cite{Gonz}.

The path integral becomes somewhat more complicated if the Laplacians
in \rref{b20} have zero-modes.  This will be the case, for example, if
$M$ admits more than one flat connection, as it always does for the
topologies we have considered so far.  In that case, the partition
function will involve an integral of the torsion \rref{b20} over the
moduli space ${\cal M}/{\cal D}$ of flat connections, and will typically
acquire infrared divergences.  Addition complications arise for spacetimes
with boundaries, but the results may still be expressed in terms of
standard topological invariants.  The full path integral in is considered
in all its glory in \cite{CarCos}.

To relate these path integral expressions to the quantizations considered
in the subsections \ref{RPS}--\ref{CCQ}, we should consider the Ray-Singer
torsion for a manifold with the topology $[0,1]\!\times\!\Sigma$, with
specified flat connections at the two boundaries $\{0\}\!\times\!\Sigma$
and $\{1\}\!\times\!\Sigma$.  It is not hard to show that the result is
simply a delta function that equates the initial and final holonomies.
The path integral thus reproduces the ``frozen time'' of Chern-Simons
quantization \cite{Gonz}.

We shall see in the next section that more interesting path integrals
arise when one considers topology change.  Euclidean path integrals
have also been used to explore the effects of spacetime topology on
the Hartle-Hawking wave function and the partition function in
(2+1)-dimensional quantum cosmology \cite{Fuji1,Fuji2,Fuji3,entropy,
sums,wormholes,Mizo2}, and Martinec \cite{Mart} and Mazur \cite{Mazur}
have investigated canonical path integrals in two rather different
metric formalisms.

\end{enumerate}

\section{Assorted Topics \label{Assort}}
\setcounter{footnote}{0}

The previous section gave a brief overview of some of the approaches to
quantum gravity in 2+1 dimensions.  In this final section, I will briefly
discuss two applications of these results: the study of black hole
thermodynamics and the investigation of topology-changing amplitudes.

\subsection{Black Hole Thermodynamics \label{Therm}}

Since Hawking's discovery that quantum mechanical processes cause
black holes to radiate \cite{Hawking}, black hole thermodynamics
has become a natural testing ground for quantum gravity.  It is
therefore of interest to apply the methods of the last section to
the (2+1)-dimensional black hole.  Unfortunately, we immediately run
into difficulties: as noted at the end of subsection \ref{BH}, the
parameters $r_\pm$ that determine the geometric structure and the
Chern-Simons holonomies for the black hole have no obvious canonical
conjugates, and there seems to be no classical Poisson algebra to
quantize.  This problem arises because the black hole spacetime is not
compact; the ordinary Einstein-Hilbert action must be supplemented by
boundary terms at infinity and (as a careful analysis shows \cite{CarTeit,
BTZ2,LouWhit}) at the horizon as well.  These boundary terms contain
additional parameters that serve as conjugate variables to the $r_\pm$.

We can avoid worrying about these boundary terms, however, by making
a clever choice of our method of quantization.  Rather than using
Schwarzschild time $t$ as our time coordinate, let us quantize on
surfaces of constant $r$, as is common in conformal field theory
\cite{Banksa}.  As long as we stay away from the horizons $r=r_\pm$
and from spatial infinity, we should not have to deal with boundary
terms in such a ``radial quantization.''

Let us therefore a consider cylindrical slice of constant $r$ connecting
a circle at initial time $t_1$ and one at final time $t_2$.  The geometry
of such a slice can no longer be described by the holonomy in the
angular direction alone; the Chern-Simons connections $A^{(\pm)}$ now
describe parallel transport along open segments between $t_1$ and $t_2$
as well.  The holonomy along a segment $\delta$ starting at $(t_1,\phi_1)$
and ending at $(t_2,\phi_2)$ will be characterized by two new
parameters,\footnote{The flatness of the connection guarantees that
this holonomy is unchanged if $\delta$ is smoothly deformed, as long as
its endpoints are held fixed, so these are the only new parameters.}
$\Delta\phi$ and $\Delta t$.  The segment $\delta$ is linked with the
closed curve $\gamma$ of subsection \ref{BH}, and the Poisson brackets
among $r_+$, $r_-$, $\Delta\phi$, and $\Delta t$ are consequently
nontrivial.

These brackets have been analyzed for the Euclidean black hole in
reference \cite{CarTeit}, where it is shown that they are equivalent
to brackets among metric parameters in a minisuperspace model.  The
Euclidean black hole---which is the configuration relevant for path
integral computations of thermodynamic behavior---has constant negative
curvature in three dimensions.  It is therefore locally isometric to
hyperbolic three-space $\IH^3$ with the Poincar\'e metric
\beq
ds^2 = {\ell^2\over z^2} (dx^2 + dy^2 + dz^2) , \quad z>0 ,
\label{c1}
\eeq
the Euclidean counterpart of \rref{a53}.  The identifications corresponding
to \rref{a54} now take the form
\beq
(R,\theta,\chi) \sim
  (Re^{2\pi r_+/\ell} , \theta + {2\pi|r_-|\over\ell}, \chi)
\label{c2}
\eeq
where I have expressed the ``Cartesian'' coordinates $(x,y,z)$ for
the upper half-plane in terms of ``spherical'' coordinates $(R,\theta,
\chi)$:
\beq
(x,y,z) = (R\cos\theta\cos\chi,R\sin\theta\cos\chi,R\sin\chi) .
\label{c2a}
\eeq
Physically, $\ln R$ is an angular coordinate, proportional to
the Schwarzschild angle $\phi$; the azimuthal angle $\theta$ of the
upper half-plane description is a time coordinate; and $\chi$ measures
radial distance.  A fundamental region for the identifications \rref{c2}
is shown in figure 4.
\begin{figure}
\begin{center}
\leavevmode
\epsfysize=3in \epsfbox{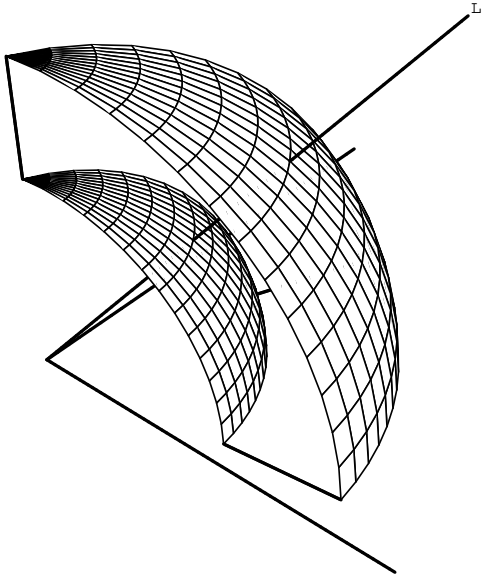}
\end{center}
\vspace{-2ex}
\caption{\small A fundamental region for the Eucidean black hole in the
upper half-space representation is obtained by identifying the inner and
outer hemispheres along radial lines such as $L$.}
\end{figure}

For the Euclidean black hole, the $\hbox{SL}(2,\IR)\!\times\!\hbox{SL}
(2,\IR)$ holonomy of equation \rref{a55} is analytically continued to
become an $\hbox{SL}(2,\IC)$ holonomy,
\beq
\rho[\gamma] = \left( \begin{array}{cc}
 e^{\pi (r_+ + i|r_-|)/\ell} & 0 \\ 0 & e^{-\pi (r_+ + i|r_-|)/\ell}
 \end{array}\right) .
\label{c3}
\eeq
On the other hand, if we consider a segment $\delta$ connecting the
points $(R_1,\theta_1)$ and $(R_2 = e^\Sigma R_1, \theta_2 = \theta_1
+ \Theta)$ along a surface of constant ``radius'' $\chi$, we find a
holonomy
\beq
\rho[\delta] = \left( \begin{array}{cc}
 e^{\pi(\Sigma + i\Theta)/\ell} & 0 \\ 0 & e^{-\pi(\Sigma + i\Theta)/\ell}
 \end{array}\right) .
\label{c4}
\eeq
The Poisson brackets of the Chern-Simons theory then give brackets
\begin{eqnarray}
\left\{r_+, \Theta\right\} &=& \left\{ |r_-|, \Sigma\right\} =
   4G \nonumber\\
\left\{ |r_-|, \Theta\right\} &=& \left\{r_+, \Sigma\right\} = 0 ,
\label{c5}
\end{eqnarray}
providing us with a starting point for quantization.

What is the significance of the new parameters $\Theta$ and $\Sigma$?
Recall that the Euclidean black hole metric is periodic in imaginary
time, and that this periodicity determines the Hawking temperature.
In the upper half-plane coordinates of \rref{c1}, imaginary Schwarzschild
time becomes to the ``spherical'' coordinate $\theta$, and periodicity
requires identifying the two endpoints of the segment $\delta$.  The
resulting metric may be shown to be smooth at the horizon only if
$\Sigma\!=\!0$ and $\Theta\!=\!2\pi$.\footnote{The resulting Hawking
temperature is easily calculated; one obtains $\beta = {2\pi r_+\ell^2
\over r_+^2-r_-^2} .$}  Other values of these parameters correspond to
Euclidean black holes with conical singularities at the horizon; as
shown in \cite{CarTeit}, $2\pi -\Theta$ is a deficit angle, while
$\Sigma$ represents a ``helical twist.''  Historically, this analysis
of the three-dimensional black hole was the starting point for the
observation \cite{CarTeit2} that conical singularities at the horizon
may play an important role in black hole thermodynamics in arbitrary
dimensions.

There is another, potentially even more powerful, way to employ the
(2+1)-dimensional black hole to gain insight into black hole
thermodynamics.  In the Euclidean model discussed above, the horizon
shrinks to a circle ($x\!=\!y\!=0$ in the coordinates \rref{c1}),
and the conical singularity must encode its full dynamics.  One may
instead begin with the Lorentzian black hole and search directly for
horizon dynamics that might provide a ``microscopic'' basis for black
hole thermodynamics.

At first sight, this seems an unlikely prospect:  as I have stressed
throughout these lectures, (2+1)-dimensional gravity is characterized
by a small number of global degrees of freedom, and there seems to be
no room for ``statistical mechanics.''  Chern-Simons theories have a
very peculiar feature, however: a Chern-Simons theory on a manifold
$M$ with boundary can induce a dynamical Wess-Zumino-Witten theory,
with an infinite number of degrees of freedom, on the boundary
$\partial M$ \cite{MS,EMSS}.  This phenomenon occurs because the presence
of a boundary breaks the gauge symmetry, allowing would-be ``pure
gauge'' degrees of freedom to become dynamical.  It may therefore make
sense to look at the event horizon of a black hole as a boundary, as
suggested by the idea of ``black hole complementarity'' \cite{Suss,Mag},
and investigate the induced WZW theory.

This program was begun in reference \cite{SC3}.  Starting with the
Chern-Simons action \rref{a18} and boundary conditions corresponding
to the presence of an apparent horizon, it is easy to find the
induced $\hbox{SL}(2,\IR)\!\times\!\hbox{SL}(2,\IR)$ WZW action.
This model is not completely understood, but in the large $k$---i.e.,
small $\Lambda$---limit, it may be approximated as a theory of six
independent bosonic string oscillators.  Such a system has an infinite
number of states, but a remaining gauge invariance reduces these to
a finite number of physical states.  With plausible (although by no
means certain) assumptions about the relevant boundary conditions and
representations of operator algebras, these states can be counted;
their number behaves asymptotically as
\beq
\log n(r_+) \sim {2\pi r_+\over 4G} ,
\label{c6}
\eeq
giving precisely the right expression for the entropy of the
(2+1)-dimensional black hole \cite{BTZ,CarTeit}.  Important
uncertainties remain, but (2+1)-dimensional gravity may be offering
a first glimpse into the quantum gravitational dynamics responsible
for black hole thermodynamics.

\subsection{Topology-Changing Amplitudes \label{Top}}

Let me conclude with one more topic, the question of whether quantum
gravity allows changes in the topology of space.  Classically, it may
be shown that topology-changing processes are forbidden \cite{Geroch},
but there has long been speculation that quantum mechanical tunneling
between spatial topologies may be allowed.

The natural way to approach such a question is by means of a path
integral \cite{Witb}.  As discussed in subsection \ref{Other}, path
integrals in (2+1)-dimensional gravity may be computed exactly, yielding
topological invariants such as Reidemeister torsion.  To compute
a topology-changing amplitude, we choose a manifold $M$ whose boundary
\beq
\partial M = \Sigma_1 \amalg \Sigma_2
\label{c7}
\eeq
is the disjoint union of an ``initial'' surface $\Sigma_1$ and a ``final''
surface $\Sigma_2$, and we specify an induced spin connection $\tilde\omega$
on $\partial M$ as boundary data.  We must then consider path integrals of
the form
\beq
Z_M[\tilde\omega] = \int [d\omega][de]\,
 \exp\left\{ iI_{\hbox{\scriptsize grav}}[M] \right\} ,
\label{c8}
\eeq
where $I_{\hbox{\scriptsize grav}}$ is the first-order action \rref{a8}.
Amano and Higuchi have shown that the requirement that the boundaries
of $M$ be spacelike leads to ``topological selection rules,'' essentially
requiring that the Euler characteristics of the initial and final spatial
slices be equal \cite{Amano}.  These rules prevent a transition from a
connected surface of genus $g$ to one of genus $g'\!\ne\!g$.  Since
$\Sigma_1$ and $\Sigma_2$ may each have many connected components,
however, this restriction still leaves room for a good deal of topology
change.

The evaluation of the path integral \rref{c8} requires a careful analysis
of the zero-modes of $e$ and $\omega$, whose number depends delicately
on the boundary conditions.  Such an analysis has been carried out in
reference \cite{CarCos}.  As Witten first pointed out, one will typically
find infrared divergences coming from integrals over zero-modes of $e$;
these essentially correspond to integrals over arbitrary lengths of closed
curves in $M$, that is, ``large universes.''  Up to the problem of
regulating these divergences, however, the computation may be carried
out explicitly for simple topologies.  Reference \cite{CarCos}, for
example, includes a computation of the amplitude for a genus 3 space
to split into two disconnected genus 2 pieces.  The possibility of
performing such an exact computation is certainly peculiar to 2+1
dimensions, but this simple model at least shows that topology change
is not forbidden by any fundamental principles.  On the other hand,
canonical quantum gravity on a manifold $\IR\!\times\!\Sigma$, with no
topology change, is also consistent, so it appears that some choices
are available.

\section{Conclusion}

Spacetime is not three-dimensional, and (2+1)-dimensional gravity is
clearly not a physically realistic model of our universe.  Nevertheless,
I hope I have convinced you that this simple model is rich enough to
allow us to learn a good deal about the nature of quantum gravity.
In particular, the method of covariant canonical quantization, with
the approach to observables described in subsection \ref{CCQ}, seems
to offer a promising approach to some of the basic conceptual problems
of quantum gravity.  Moreover, the analyses of black hole thermodynamics
and topology change may offer genuine physical insight into the real
(3+1)-dimensional world.

Needless to say, there remains a great deal to be learned from this
model.  Of special interest is the problem of coupling matter to quantum
gravity.  There is an old suggestion that nonperturbative quantum
gravitational effects might cut off the divergences of ordinary quantum
field theories \cite{DeWitta,Ish1,Ish2}.  Ashtekar and Varadarajan have
recently found evidence for this conjecture in 2+1 dimensions, in the
form of an observation that the Hamiltonian of a scalar field coupled
to (2+1)-dimensional gravity is bounded {\em above} \cite{AshVar}.
Another interesting observation, due to Witten \cite{Witd,Becker}, is
that in the presence of gravity in 2+1 dimensions, unbroken supersymmetry
may lead to a vanishing cosmological constant {\em without\/} requiring
the equality of boson and fermion masses.  For researchers in this field,
much interesting work remains.

\vspace{1.5ex}
\begin{flushleft}
\large\bf Acknowledgements
\end{flushleft}

This work was supported in part by National Science Foundation grant
PHY-93-57203 and Department of Energy grant DE-FG03-91ER40674.


\newpage
\begin{thebibliography}{999}
\bibitem{Staru} A.\ Staruszkiewicz, {\sl Acta Phys.\ Polon.} {\bf 24}
 (1963) 734.
\bibitem{Leut} H.\ Leutwyler, Nuovo Cimento {\bf 42A} (1966) 159.
\bibitem{Collas} P.\ Collas, {\sl Am.\ J.~Phys.} {\bf 45} (1977) 833.
\bibitem{Clem1} G.\ Cl\'ement, \NPB{114} (1976) 437.
\bibitem{DJtH} S.\ Deser, R.\ Jackiw, and G.\ 't~Hooft, \Ann{152} (1984) 220.
\bibitem{DJ} S.\ Deser and R.\ Jackiw, \CMP{118} (1988) 495.
\bibitem{tH} G.\ 't~Hooft, \CMP{117} (1988) 685.
\bibitem{Wita} E.\ Witten, \NPB{311} (1988) 46.
\bibitem{Witb} E.\ Witten, \NPB{323} (1989) 113.
\bibitem{Witc} E.\ Witten, \CMP{137} (1991) 29.
\bibitem{Achu} A.\ Ach\'ucarro and P.~K.\ Townsend, \PLB{180} (1986) 89.
\bibitem{Gott} J.~R.\ Gott, \PRL{66} (1991) 1126.
\bibitem{Gott2} S.~M.\ Carroll et al., \PRD{50} (1994) 6190.
\bibitem{tH2} G.\ 't~Hooft, \CQG{9} (1992) 1335.
\bibitem{Mena} P.\ Menotti and D.\ Seminara, \NPB{419} (1994) 189.
\bibitem{Jackiw} R.\ Jackiw, in {\sl Proc.\ of the Sixth Marcel Grossmann
 Meeting on General Relativity}, edited by H.\ Sato and T.\ Nakamura
 (World Scientific, Singapore, 1992).
\bibitem{Rocek} M.\ Ro{\v c}ek and R.~M.\ Williams, \CQG{2} (1985) 701.
\bibitem{Clem} G.\ Cl\'ement, \IJTP{24} (1985) 267.
\bibitem{Gerbert} P.\ Gerbert and R.\ Jackiw, \CMP{124} (1989) 229.
\bibitem{SC1} S.\ Carlip, \NPB{324} (1989) 106.
\bibitem{Koehler} K.\ Koehler et al., \NPB{348} (1991) 373.
\bibitem{SCm} S.\ Carlip, \CQG{8} (1991) 5.
\bibitem{Vaz} C.\ Vaz and L.\ Witten, \MPLA{7} (1992) 2763.
\bibitem{Mansouri} M.~K.\ Falbo-Kenkel and F.\ Mansouri, \JMP{34} (1993) 139.
\bibitem{Ortiz} D.\ Kabat and M.~E.\ Ortiz, \PRD{49} (1994) 1684.
\bibitem{Geg} J.\ Gegenberg, G.\ Kunstatter, and H.~P.\ Leivo, \PLB{252}
 (1990) 381.
\bibitem{CarGeg} S.\ Carlip and J.\ Gegenberg, \PRD{44} (1991) 424.
\bibitem{Mizo} S.\ Mizoguchi and H.\ Yamamoto, \PRD{50} (1994) 7351.
\bibitem{AshVar} A.\ Ashtekar and M.\ Varadarajan, \PRD{50} (1994) 4944.
\bibitem{Pel} P.\ Peldan, \NPB{395} (1993) 239.
\bibitem{Bona} G.\ Bonacina, A.\ Gamba, and M.\ Martellini, \PRD{45}
 (1992) 3577.
\bibitem{Dayi} O.~F.\ Dayi, \PLB{234} (1990) 25.
\bibitem{Ko2} K.\ Koehler et al., \JMP{32} (1991) 239.
\bibitem{deWit} D.\ de Wit, H.~J.\ Matschull, and H.\ Nicolai, \PLB{318}
 (1993) 115.
\bibitem{Nicolai} H.~J.\ Matschull and H.\ Nicolai, \NPB{411} (1994) 609.
\bibitem{Henn} M.\ Henneaux, \PRD{29} (1984) 2766.
\bibitem{Deser} S.\ Deser, \CQG{2} (1985) 489.
\bibitem{Brown} J.~D.\ Brown, {\sl Lower Dimensional Gravity} (World
 Scientific, Singapore, 1988).
\bibitem{Menotti} P.\ Menotti and D.\ Seminara, \Ann{208} (1991) 449.
\bibitem{Menotti2} P.\ Menotti and D.\ Seminara, \NPB{376} (1992) 411.
\bibitem{DJTemp} S.\ Deser, R.\ Jackiw, and S.\ Templeton, \Ann{140}
 (1982) 372.
\bibitem{DesY} S.\ Deser and Z.\ Yang, \CQG{7} (1990) 1603.
\bibitem{Kleppe} B.\ Keszthelyi and G.\ Kleppe, \PLB{281} (1992) 33.
\bibitem{Grig1} G.\ Grignani and G.\ Nardelli, \PLB{264} (1991) 45.
\bibitem{Grig2} G.\ Grignani and G.\ Nardelli, \NPB{370} (1992) 491.
\bibitem{Kawai} T.\ Kawai, \PRD{49} (1994) 2862.
\bibitem{Brooks} R.\ Brooks, \MPLA{8} (1993) 2277.
\bibitem{Brooks2} R.\ Brooks and G.\ Lifschytz, ``Quantum Gravity and
 Equivariant Cohomology,'' MIT preprint MIT-CTP-2340 (1994).
\bibitem{Abikoff} W.\ Abikoff, {\sl The Real Analytic Theory of
 Teichm\"uller Space}, Lecture Notes in Mathematics {\bf 820}
 (Springer-Verlag, Berlin, 1980).
\bibitem{Fish} A.\ Fischer and A.\ Tromba, {\sl Math.\ Ann.} {\bf 267}
 (1984) 311.
\bibitem{Keen} For a nice introduction, see L.\ Keen, {\sl Math.\
 Intelligencer} {\bf 16} (1994) 11.
\bibitem{Thur} W.~P.\ Thurston, {\sl The Geometry and Topology of
 Three-Manifolds}, Princeton lecture notes (1979).
\bibitem{Canary}  R.~D.\ Canary, D.~B.~A.\ Epstein, and P.\ Green,
 in {\sl Analytical and Geometric Aspects of Hyperbolic Space},
 London Math.~Soc.~Lecture Notes Series {\bf 111}, edited by D.~B.\ Epstein
 (Cambridge University Press, Cambridge, 1987).
\bibitem{Gold3} W.~M.\ Goldman, in {\sl Geometry of Group Representations},
 edited by W.~M.\ Goldman and A.~R.\ Magid (American Mathematical Society,
 Providence, 1988).
\bibitem{SullThur} For examples of geometric structures, see
 D.\ Sullivan and W.\ Thurston, {\sl Enseign.\ Math.} {\bf 29} (1983) 15.
\bibitem{SC2} S.\ Carlip, in {\sl Knots and Quantum Gravity},
 edited by J.\ Baez (Clarendon Press, Oxford, 1994).
\bibitem{Mess} G.\ Mess, ``Lorentz Spacetimes of Constant Curvature,''
 Institut des Hautes Etudes Scientifiques preprint IHES/M/90/28 (1990).
\bibitem{Fujiwara} Y.\ Fujiwara, \CQG{10} (1993) 219.
\bibitem{Ezawa} K.\ Ezawa, \PRD{49} (1994) 5211; addendum in
 \PRD{50} (1994) 2935.
\bibitem{Rom} J.~D.\ Romano, \GRG{25} (1993) 759.
\bibitem{Ban} M.\ Ba\~nados, ``Sugawara Construction and the Relation
 between Diffeomorphisms and Gauge Transformations in General Relativity,''
 Imperial College preprint Imperial-TP-93-94-40 (1994).
\bibitem{WitJones} E.\ Witten, \CMP{121} (1989) 351.
\bibitem{NR1} J.~E.\ Nelson and T.\ Regge, \NPB{328} (1989) 190.
\bibitem{NR3} J.~E.\ Nelson and T.\ Regge, \CQG{9} (1992) 187.
\bibitem{NR2} J.~E.\ Nelson and T.\ Regge, Phys.\ Lett.\ {\bf B272}
 (1991) 213.
\bibitem{Martin} S.\ Martin, \NPB{327} (1989) 178.
\bibitem{Loll} R.\ Loll, ``Wilson Loop Coordinates for 2+1 Gravity,''
 Penn State preprint CGPG-94-8-1 (1994).
\bibitem{NRZ} J.~E.\ Nelson, T.\ Regge, and F.\ Zertuche, \NPB{339}
 (1990) 516.
\bibitem{SC3} S.\ Carlip, \PRD{51} (1995) 632.
\bibitem{Goldman4} W.~M.\ Goldman, {\sl Adv.\ Math.} {\bf 54} (1984) 200.
\bibitem{GoldHam} W.~M.\ Goldman, {\sl Invent. Math.} {\bf 85} (1986) 263.
\bibitem{Mon} V.\ Moncrief, \JMP{30} (1989) 2907.
\bibitem{HosNak} A.\ Hosoya and K.\ Nakao, \CQG{7} (1990) 163.
\bibitem{York} J.~W.\ York, \PRL{28} (1972) 1082.
\bibitem{Macbeath} A.~M.\ Macbeath and D.\ Singerman, {\sl Proc.\ London
 Math.\ Soc.} {\bf 31} (1975) 211.
\bibitem{Goldman2} W.~M.\ Goldman, in {\sl Geometry and Topology}, Lecture
 Notes in Mathematics 1167, edited by J.\ Alexander and J.\ Harer (Springer,
 Berlin, 1985).
\bibitem{Goldman3} W.~M.\ Goldman, {\sl Invent.\ Math.} {\bf 93} (1988)
 557.
\bibitem{Birman} J.~S.\ Birman and H.~M.\ Hilden, in {\sl Advances in the
 Theory of  Riemann Surfaces}, Annals of Math.~Studies {\bf 66}, edited
 by L.~V.\ Ahlfors  {et al.} (Princeton University Press, Princeton, 1971).
\bibitem{Birman2} J.~S.\ Birman, in {\sl Discrete Groups and Automorphic
 Functions}, edited by W.~J.\ Harvey (Academic Press, New York, 1977).
\bibitem{CarNel} S.\ Carlip and J.~E.\ Nelson, ``Comparative Quantizations
 of (2+1)-Dimensional Gravity,'' Davis preprint UCD-94-37 and Torino preprint
 DFTT/49/94 (1994), to appear in {\sl Phys.~Rev.} {\bf D}.
\bibitem{CarNel1} S.\ Carlip and J.~E.\ Nelson, \PLB{324} (1994) 299.
\bibitem{Ezawa2} K.\ Ezawa, ``Chern-Simons Quantization of (2+1) Anti-de
 Sitter Gravity on a Torus,'' Osaka preprint OU-HET/201 (1994).
\bibitem{LouMar} J.\ Louko and D.~M.\ Marolf, \CQG{11} (1994) 311.
\bibitem{SCobserv} S.\ Carlip, \PRD{42} (1990) 2647.
\bibitem{Soda} Y.\ Fujiwara and J.\ Soda, \PTP{83} (1990) 733.
\bibitem{BTZ} M.\ Ba\~nados, C.\ Teitelboim, and J.\ Zanelli, \PRL{69}
 (1992) 1849.
\bibitem{BHTZ} M.\ Ba\~nados, M.\ Henneaux, C.\ Teitelboim, and J.\
 Zanelli, \PRD{48} (1993) 1506.
\bibitem{CarHos} S.\ Carlip and A.\ Hosoya, in preparation.
\bibitem{BrownMann} J.~D.\ Brown, J.\ Creighton and R.~B.\ Mann,
 \PRD{50} (1994) 6394.
\bibitem{CarTeit} S.\ Carlip and C.\ Teitelboim, \PRD{51} (1995) 622.
\bibitem{Manna} D.\ Cangemi, M.\ Leblanc, and R.~B.\ Mann, \PRD{48}
 (1993) 3606.
\bibitem{HosNak2} A.\ Hosoya and K.\ Nakao, \PTP{84} (1990) 739.
\bibitem{Puzio} R.~S.\ Puzio, \CQG{11} (1994) 2667.
\bibitem{Maass} See, for example, H.\ Iwaniec, in {\sl Modular Forms},
 edited by R.~A.\ Rankin (Ellis Horwood Ltd., Chichester, 1984).
\bibitem{Puzio2} R.\ Puzio, \CQG{11} (1994) 609.
\bibitem{Maass2} J.~D.\ Fay, {\sl J.~Reine\ Angew.\ Math.} {\bf 293}
 (1977) 143.
\bibitem{Maass3} H.\ Maass, {\sl Lectures on Modular Functions of One
 Complex Variable} (Tata Institute, Bombay, 1964).
\bibitem{dirac} S.\ Carlip, \PRD{45} (1992) 3584.
\bibitem{ordering} S.\ Carlip, \PRD{47} (1993) 4520.
\bibitem{Isen} J.\ Isenberg and J.~E.\ Marsden, {\sl Phys.\ Reports}
 {\bf 89} (1982) 179.
\bibitem{Kuchar} K.\ Kucha{\v r}, in {\sl Proc.\ of the 4th Canadian
 Conf.\ on General Relativity and Relativistic Astrophysics}, edited by G.\
 Kunstatter et al.\ (World Scientific, Singapore, 1992).
\bibitem{NR4} J.~E.\ Nelson and T.\ Regge, \CMP{141} (1991) 211.
\bibitem{NR5} J.~E.\ Nelson and T.\ Regge, \CMP{155} (1993) 561.
\bibitem{NR6} J.~E.\ Nelson and T.\ Regge, \PRD{50} (1994) 5125.
\bibitem{Rovelli1} C.\ Rovelli, \PRD{42} (1990) 2638.
\bibitem{Rovelli2} C.\ Rovelli, \PRD{43} (1991) 442.
\bibitem{AshMag} A.\ Ashtekar and A.\ Magnon, {\sl Proc.\ Roy.\ Soc.\
 (London)} {\bf A346} (1975) 375.
\bibitem{Crn} C.\ Crnkovic and E.\ Witten, in {\sl Three Hundred
 Years of Gravity}, edited by S.~W.\ Hawking and W.\ Israel (Cambridge
 University Press, Cambridge, 1987).
\bibitem{AshBom} A.\ Ashtekar. L.\ Bombelli, and O.\ Reula, in {\sl
 Mechanics, Analysis and Geometry: 200 Years after Lagrange}, edited
 by M.\ Francaviglia (Elsevier Science Publishers, Amsterdam, 1991).
\bibitem{Wald} R.~M.\ Wald, {\sl Quantum Field Theory in Curved Spacetime
 and Black Hole Thermodynamics} (University of Chicago Press, Chicago,
 1994).
\bibitem{Isham} See, for example, C.~J.\ Isham, in {\sl Quantum Gravity},
 edited by C.~J.\ Isham, R.\ Penrose, and D.~W.\ Sciama (Clarendon Press,
 Oxford, 1975).
\bibitem{Ezawa3} K.\ Ezawa, \IJMPA{9} (1994) 4727.
\bibitem{Peldan} P.\ Peldan, ``Large Diffeomorphisms in (2+1) Quantum
 Gravity on the Torus,'' Penn State preprint CGPG-95-1-1 (1995).
\bibitem{Six} S.\ Carlip, in {\sl Proc.\ of the  Fifth Canadian Conference
 on General Relativity and Relativistic Astrophysics}, edited by R.~B.\ Mann
 and R.~G.\ McLenaghan (World Scientific, Singapore, 1994).
\bibitem{WdW} S.\ Carlip, \CQG{11} (1994) 31.
\bibitem{Henn2} M.\ Henneaux, \PLB{134} (1984) 184.
\bibitem{Visser} M.\ Visser, \PRD{42} (1990) 1964.
\bibitem{Banks} T.\ Banks, W.\ Fischler, and L.\ Susskind, \NPB{262}
 (1985) 159.
\bibitem{Ashbook} A.\ Ashtekar, {\sl Lectures on Nonperturbative Quantum
 Gravity} (World Scientific, Singapore, 1991).
\bibitem{Ash2} A.\ Ashtekar et al., \CQG{6} (1989) L185.
\bibitem{RovSmol} C.\ Rovelli and L.\ Smolin, \NPB{331} (1990) 80.
\bibitem{Marolf} D.\ Marolf, \CQG{10} (1993) 2625.
\bibitem{Smolin} L.\ Smolin, in {\sl Knots, Topology, and Quantum
 Field Theories}, edited by L.\ Lusanna (World Scientific, Singapore,
 1989).
\bibitem{AshLoll} A.\ Ashtekar and R.\ Loll, \CQG{11} (1994) 2417.
\bibitem{Wael} H.\ Waelbroeck, \CQG{7} (1990) 751.
\bibitem{waela} H.\ Waelbroeck, \PRL{64} (1990) 2222.
\bibitem{waelb} H.\ Waelbroeck and F.\ Zertuche, \PRD{50} (1994) 4966.
\bibitem{tH3} G.\ 't~Hooft, \CQG{10} (1993) S79.
\bibitem{NewUn} W.~G.\ Unruh and P.\ Newbury, \IJMPD{3} (1994) 131.
\bibitem{PonReg} G.\ Ponzano and T.\ Regge, in {\sl Spectroscopic and
 Group Theoretical Methods in Physics}, edited by F.\ Block (North-Holland,
 Amsterdam, 1968).
\bibitem{Hass} B.\ Hasslacher and M.~J.\ Perry, \PLB{103} (1981) 21.
\bibitem{TurVir} V.~G.\ Turaev and O.~Y.\ Viro, {\sl Topology} {\bf 31}
 (1992) 865.
\bibitem{Oog1} H.\ Ooguri and N.\ Sasakura, \MPLA{6} (1991) 3591.
\bibitem{Oog2} H.\ Ooguri, \NPB{382} (1992) 276.
\bibitem{ArchWill}  F.\ Archer and R.~M.\ Williams, \PLB{273} (1991) 438.
\bibitem{Rovlat} C.\ Rovelli, \PRD{48} (1993) 2702.
\bibitem{Iwa} J.\ Iwasaki, ``A Reformulation of the Ponzano-Regge Quantum
 Gravity Model in Terms of Surfaces,'' Pittsburgh preprint (1994).
\bibitem{Fox} T.~J.\ Foxon, ``Spin Networks, Turaev-Viro Theory and the
 Loop Representation,'' Cambridge preprint DAMTP-R-94-30 (1994).
\bibitem{Barrett} J.~W.\ Barrett and T.~J.\ Foxon, \CQG{11} (1994) 543.
\bibitem{RaySinger} D.~B.\ Ray and I.~M.\ Singer, {\sl Adv.\ Math.}
 {\bf 7} (1971) 145.
\bibitem{Milnor} J.~W.\ Milnor, {\sl Bull.\ Amer.\ Math.\ Soc.} {\bf 72}
 (1966) 358.
\bibitem{Luck} W.\ L{\"u}ck, {\sl J.~Diff.\ Geom.} {\bf 37} (1993) 263.
\bibitem{GegKun} J.\ Gegenberg and G.\ Kunstatter, \Ann{231} (1994) 270.
\bibitem{Gonz} G.\ Gonzalez and J.\ Pullin, \PRD{42} (1990) 3395.
\bibitem{CarCos} S.\ Carlip and R.\ Cosgrove, \JMP{35} (1994) 5477.
\bibitem{Fuji1} Y.\ Fujiwara et al., \PRD{44} (1991) 1756.
\bibitem{Fuji2} Y.\ Fujiwara et al., \PRD{44} (1991) 1763.
\bibitem{Fuji3} Y.\ Fujiwara et al., \CQG{9} (1992) 867.
\bibitem{entropy} S.\ Carlip, \PRD{46} (1992) 4387.
\bibitem{sums} S.\ Carlip, \CQG{10} (1993) 207.
\bibitem{wormholes} S.\ Carlip and S.~P.\ de~Alwis, \NPB{337} (1990) 681.
\bibitem{Mizo2} S.\ Mizoguchi, \MPLA{8} (1993) 3909.
\bibitem{Mart} E.\ Martinec, \PRD{30} (1984) 1198.
\bibitem{Mazur} P.~O.\ Mazur, \PLB{262} (1991) 405.
\bibitem{Hawking} S.~W.\ Hawking, {\sl Nature} {\bf 248} (1974) 30.
\bibitem{BTZ2} M.\ Ba\~nados, C.\ Teitelboim, and J.\ Zanelli, \PRL{72}
 (1994) 957.
\bibitem{LouWhit} J.\ Louko and B.~F.\ Whiting, ``Hamiltonian Thermodynamics
 of the Schwarzschild Black  Hole,'' Florida preprint UF-RAP-94-13 (1994).
\bibitem{Banksa} See, for example, T.\ Banks, in {\sl The Santa Fe TASI-87},
 edited by R.\ Slansky and G.\ West (World Scientific, Singapore, 1988).
\bibitem{CarTeit2} S.\ Carlip and C.\ Teitelboim, ``The Off-Shell Black
 Hole,'' IAS preprint IASSNS-HEP-93/84 and Davis preprint UCD-93-34 (1993).
\bibitem{MS} G.\ Moore and N.\ Seiberg, \PLB{220} (1989) 422.
\bibitem{EMSS} S.\ Elitzur et al., \NPB{326} (1989) 108.
\bibitem{Suss} L.\ Susskind, L.\ Thorlacius, and J.\ Uglum, \PRD{48}
 (1993) 3743.
\bibitem{Mag} M.\ Maggiore, \PLB{333} (1994) 39.
\bibitem{Geroch} R.~P.\ Geroch, \JMP{8} (1967) 782.
\bibitem{Amano} K.\ Amano and S.\ Higuchi, \NPB{377} (1992) 218.
\bibitem{DeWitta} B.~S.\ DeWitt, \PRL{13} (1964) 114.
\bibitem{Ish1} C.~J.\ Isham, A.\ Salam, and J.\ Strathdee, \PRD{3}
 (1971) 1805.
\bibitem{Ish2} C.~J.\ Isham, A.\ Salam, and J.\ Strathdee, \PRD{5}
 (1972) 2548.
\bibitem{Witd} E.\ Witten, ``Is Supersymmetry Really Broken?'' IAS preprint
 IASSNS-HEP-94-72 (1994).
\bibitem{Becker} K.\ Becker, M.\ Becker, and A.\ Strominger,
 ``Three-Dimensional Supergravity and the Cosmological Constant,''
 Santa Barbara ITP preprint NSF-ITP-95-07 (1995).
\end{thebibliography}
\end{document}